\documentclass[12pt,a4paper]{article}
\pdfoutput=1

\usepackage[T1]{fontenc} 
\usepackage{microtype} 

\usepackage[hmarginratio=1:1,top=32mm,columnsep=20pt]{geometry} 

\usepackage{amsmath,amssymb}
\usepackage{slashed}
\usepackage{hyperref}
\usepackage{cite}
\usepackage{graphicx}
\usepackage[percent]{overpic}

\usepackage{abstract}

\usepackage{titlesec} 
\titleformat{\section}[block]{\large\bfseries\centering}{\thesection}{1em}{} 
\titleformat{\subsection}[block]{\bfseries}{\thesubsection}{1em}{} 
\numberwithin{equation}{section}

\usepackage{fancyhdr} 
\title{\vspace{-10mm}\fontsize{24pt}{10pt}\selectfont\textbf{Spherical Branes}\vspace{10mm}}

\author{Nikolay Bobev, Pieter Bomans and Fri{\dh}rik Freyr Gautason\\[15mm] 
\normalsize Instituut voor Theoretische Fysica, KU Leuven\\
\normalsize Celestijnenlaan 200D, B-3001 Leuven, Belgium\\[5mm]
\texttt{\small\href{mailto:nikolay.bobev@kuleuven.be}{\{nikolay.bobev}, \href{mailto:pieter.bomans@kuleuven.be}{pieter.bomans}, \href{mailto:ffg@kuleuven.be}{ffg\}@kuleuven.be}}
}
\date{}

\usepackage{amsthm}

\usepackage{enumerate}
\numberwithin{equation}{section}

\usepackage{color}
\definecolor{dark-gray}{gray}{0.20}
\definecolor{gray}{gray}{0.30}
\definecolor{light-gray}{gray}{0.80}
\definecolor{dark-red}{rgb}{0.7,0,0}
\definecolor{dark-green}{rgb}{0.1,0.4,0}
\definecolor{dark-blue}{rgb}{0.3,0.3,0.7}
\definecolor{light-blue}{rgb}{0.8,0.8,1}

\usepackage{cite}
\usepackage{hyperref}
\hypersetup{
	colorlinks=true,
	linkcolor=dark-blue,
	citecolor=dark-red,
	urlcolor=dark-green,
	linktoc=page
}

\newcommand{\dd}{\mathrm{d}}

\newcommand{\e}{\mathrm{e}}
\newcommand{\w}{\wedge}

\newcommand{\be}{\begin{equation}}
\newcommand{\ee}{\end{equation}}
\newcommand{\bea}{\begin{eqnarray}}
\newcommand{\eea}{\end{eqnarray}}

\newcommand{\f}[2]{\frac{#1}{#2}}

\newcommand{\p}[1]{\phantom{#1}}
\newcommand{\vol}{\text{vol}}

\newcommand{\Tr}{\text{Tr}~}

\newcommand{\obar}[1]{\overline{#1}}
\newcommand{\SO}{\text{SO}}
\newcommand{\SU}{\text{SU}}
\newcommand{\SL}{\text{SL}}
\newcommand{\U}{\text{U}}
\newcommand{\ISO}{\text{ISO}}

\begin{document}

\maketitle
\thispagestyle{fancy}

\bigskip
\bigskip
\bigskip

\begin{abstract}

We construct new solutions of ten-dimensional supergravity with sixteen supercharges which describe the backreaction of D$p$-branes with spherical worldvolume. These solutions are holographically dual to the $(p+1)$-dimensional maximally supersymmetric Yang-Mills theory on $S^{p+1}$. The finite size of the sphere provides an IR cut-off for the gauge theory which is manifested in the supergravity solution as a smooth cap-off of the geometry. In the UV the size of the sphere plays no role and the backgrounds asymptote to the well-known supergravity solutions that describe the near-horizon limit of flat D$p$-branes. We compute the on-shell action of our spherical brane solutions and show that it is in agreement with recent supersymmetric localization results for the free energy of maximal SYM theory on $S^{p+1}$.

\end{abstract}

\vspace{4cm}

\hspace{6cm}\textit{Dedicated to the memory of Joe Polchinski.}

\thispagestyle{empty}

\newpage

\setcounter{tocdepth}{2}
\tableofcontents

\section{Introduction}
\label{sec:Intro}

Brane solutions in supergravity have offered multiple important insights into the structure of string theory, supergravity, and holography. They were first constructed as extremal black branes in ten- and eleven-dimensional supergravity which preserve half of the maximal supersymmetry \cite{Horowitz:1991cd}. An important insight came from the realization that they should be thought of as sourced by the D- and M-branes of string and M-theory \cite{Polchinski:1995mt}, based on the earlier work \cite{Dai:1989ua,Leigh:1989jq}. The dichotomy between the gauge theory living on the worldvolume of the D-branes and their backreacted $p$-brane solutions ultimately led to the development of the gauge/gravity duality \cite{Maldacena:1997re,Gubser:1998bc,Witten:1998qj} and to important insights into black hole physics in string theory \cite{Strominger:1996sh}.

In the standard treatment of D$p$-brane supergravity solutions the world-volume of the brane is $(p+1)$-dimensional flat space, $\mathbf{R}^{1,p}$. For general values of $p$ the near-horizon limit of the supergravity background exhibits a singularity where the running dilaton diverges. This bodes well for the holographic interpretation of these backgrounds as dual to $(p+1)$-dimensional maximally supersymmetric Yang-Mills (SYM) theory on $\mathbf{R}^{1,p}$. For general values of $p$ this theory is not conformal and it is expected that the weakly curved region of the supergravity solution is dual to the regime of strong gauge coupling while for small values of the running coupling the supergravity description is not valid and the background develops a singularity \cite{Itzhaki:1998dd}.

Given the importance of flat D$p$-branes and their supergravity description it is natural to explore the more general situation when the worldvolume of the brane is curved. Since supersymmetry offers a great deal of calculational control which often elucidates the underlying physics, one is led to look for curved supersymmetric D$p$-branes. Indeed, this question was addressed in \cite{Bershadsky:1995qy} where D$p$-branes with worldvolumes of the form $\mathbf{R}^{1,m}\times \mathcal{M}_q$, with $m+q=p$ and $\mathcal{M}_q$ a general Euclidean manifold, were studied. To preserve supersymmetry the worldvolume theory on the brane is partially topologicallly twisted on $\mathcal{M}_q$ \cite{Witten:1988ze}. This setup has a beautiful extension into the arena of holography as emphasized in \cite{Maldacena:2000mw}. However it is well-known that the topological twist is not the only way by which a supersymmetric gauged theory can be placed on a curved manifold, see for example \cite{Blau:2000xg,Pestun:2007rz,Festuccia:2011ws,Pestun:2016zxk}. A particularly simple example of a curved manifold on which a SYM theory can be placed in a supersymmetric way is offered by the sphere equipped with an Einstein metric \cite{Blau:2000xg}. Thus it is natural to ask whether this gauge theory construction admits a realization in string theory on the world-volume of spherical D$p$-branes and how to construct the supergravity solutions describing the backreaction of these branes. The goal of this work is to address this question from the point of view of supergravity and holography.

Our approach to construct the supergravity solutions describing spherical D$p$-branes is informed by the knowledge of the Lagrangian of the maximally supersymmetric Yang-Mills theory on $S^{p+1}$ for $p\leq 6$ \cite{Blau:2000xg,Minahan:2015jta}. For general values of $p\neq 3$ the maximal SYM theory  is not conformal and coupling it to the curvature of the sphere while preserving sixteen supercharges necessitates certain couplings in the Lagrangian. These couplings in turn break the R-symmetry of the SYM theory\footnote{The R-symmetry group is non-compact due the fact that the SYM theory is defined in Euclidean signature.} from $\SO(1,8-p)$ to $\SO(1,2)\times \SO(6-p)$. It is natural to assume that the world-volume theory for spherical D$p$-branes at low energies is the same as this maximal SYM on $S^{p+1}$. The symmetry breaking pattern combined with the presence of sixteen supercharges then leads to a very restrictive ansatz for the type II supergravity backgrounds describing the spherical branes. Nevertheless it is still difficult to solve the supergravity BPS equations and find the explicit solutions directly in ten dimensions. We circumvent this impasse by employing the well-known technique of reducing the ten-dimensional supergravity theory to an effective gauged supergravity in $p+2$ dimensions. The spherical brane solutions of interest are then found as supersymmetric domain walls in this gauged supergravity with non-trivial profiles for the metric as well as three scalar fields.\footnote{For $p=6$ there are only two scalar fields needed in the eight-dimensional supergravity theory.} These scalar fields are the supergravity manifestation of the running gauge coupling of SYM theory and the couplings in the Lagrangian on $S^{p+1}$ that need to be turned on to preserve supersymmetry. Working with this $p+2$-dimensional gauged supergravity we are able to construct explicitly the supersymmetric spherical domain wall solutions of interest and then use standard uplift formulae from the literature to convert them to solutions of type II supergravity.

Our spherical brane solutions exhibit some common features which are in harmony with the physics of the SYM theory. In the IR region of the geometry the solution is regular and the radial coordinate combines with the metric on $S^{p+1}$ to produce a smooth cap-off that locally looks like $\mathbf{R}^{p+2}$. This behavior reflects the fact that in the dual SYM theory the length scale associated with the sphere provides and IR cut-off for the dynamics and one cannot probe energies smaller than this scale. This type of smooth cap-off of supergravity solutions dual to non-conformal gauge theories on a sphere is a familiar feature from recent holographic studies of mass deformations of three-dimensional ABJM, four-dimensional $\mathcal{N}=4$, and five-dimensional D4-D8 SCFTs \cite{Freedman:2013ryh,Bobev:2013cja,Bobev:2016nua,Bobev:2018hbq,Gutperle:2018axv}.  In the UV region of the spherical brane solutions the background is the same as the near-horizon limit of the usual flat D$p$-brane backgrounds \cite{Horowitz:1991cd}, albeit with Euclidean worldvolume. This behavior is also in line with the dual gauge theory where at high energies the radius of the sphere should not affect the dynamics and one expects to recover the physics of SYM theory in flat space. 

Part of the motivation for constructing the supergravity solutions describing spherical D$p$-branes is to be able to make contact with recent results on supersymmetric localization for maximal SYM on $S^{p+1}$ \cite{Minahan:2015jta,Minahan:2015any}. The free energy, i.e. the logarithm of the path integral, of the $\SU(N)$ SYM theory can be computed in the large $N$ limit by taking a continuous approximation of the matrix model arising from the localization calculation. It was found in \cite{Minahan:2015any} that this large $N$ free energy scales as
\begin{equation}\label{eq:Fintro}
F\sim N^2\lambda_{\rm eff}^{\frac{p-3}{5-p}}\,,
\end{equation}
where $\lambda_\text{eff}(E) = g_\text{YM}^2 N E^{p-3}$ is the effective 't Hooft coupling of the gauge theory at energy scale $E$. Given our explicit supergravity solutions describing spherical branes it is tempting to compute this free energy holographically and compare with the supersymmetric localization result. This calculation is somewhat subtle due to the fact that the solutions of interest are not asymptotically AdS and thus one cannot rely on the standard holographic dictionary. Nevertheless it is possible to use the results in \cite{Peet:1998wn,Boonstra:1998mp,Kanitscheider:2008kd} to evaluate the on-shell action of the spherical brane solutions and reproduce the scaling of the free energy with $N$ and $\lambda_\text{eff}$ in \eqref{eq:Fintro}.

The spherical D$p$-brane solutions can also be interpreted from a different vantage point. It is standard in the context of non-conformal holography to encounter supergravity solutions that exhibit IR singularities, see for example \cite{Itzhaki:1998dd,Boonstra:1998mp,Freedman:1999gk,Girardello:1999bd,Gubser:2000nd,Klebanov:2000hb}. Whenever these singularities are physically acceptable they are interpreted in the dual field theory as arising from a free or gapped phase of the IR dynamics \cite{Gubser:2000nd}. The singularity is usually remedied by replacing the singular background by a black hole solution with the same asymptotics and a regular horizon. In the dual gauge theory this corresponds to turning on finite temperature, which in turn introduces a finite IR cut-off in the gauge theory. In the context of the flat D$p$-brane solutions this is discussed in some detail in \cite{Itzhaki:1998dd}. Our spherical brane solutions provide an alternative way to excise the singularity of flat D$p$-brane supergravity backgrounds. Due to the finite length scale introduced by the sphere one finds a smooth cap-off of the metric instead of a singularity in the IR region of the geometry. We interpret this as a gravitational manifestation of the the IR cut-off for the gauge theory on $S^{p+1}$. The difference with the more common finite temperature cut-off is that spherical branes preserve sixteen supercharges which may provide better calculational control in some circumstances. We believe that this is the unique IR cut-off compatible with the maximal number of supercharges for a non-conformal SYM theory.

We start in the next section with a review of maximally SYM theory on $S^{p+1}$. We continue in Section~\ref{sec:flatDp} with a review of the well-known flat D$p$-brane solutions. In Section~\ref{sec:SpherDp} we present the general spherical D$p$-brane solutions in a unified manner and summarize how we arrive at them by uplifting supersymmetric domain wall solutions of lower-dimensional gauged supergravity. In Section~\ref{sec:Physics} we discuss the physical interpretation of these spherical brane solutions for $1\leq p\leq 6$. Our supergravity backgrounds have a clear holographic interpretation which we discuss in Section~\ref{sec:holography}. Section~\ref{sec:conclusions} is devoted to our conclusions and a short discussion. In the three appendices we present our conventions, review the known flat Euclidean D-brane solutions of type II supergravity, and summarize the various lower-dimensional gauged supergravity theories used to construct the spherical brane solutions.

\section{SYM theory on a sphere}
\label{sec:fieldtheory}

An important guiding principle for constructing supersymmetric spherical D$p$-brane solutions is the fact that the low-energy dynamics on the worldvolume of D-branes in flat space is given by a maximally supersymmetric Yang-Mills theory. Thus it is natural to expect that for spherical D$p$-branes the low-energy physics is the same as that of maximal SYM theory on $S^{p+1}$. Since for general values of $p$,  maximal SYM is not conformal it is non-trivial to couple it to the curvature of the sphere. It is therefore useful to briefly review the construction of the Lagrangian of maximal SYM on $S^{p+1}$.

Maximally supersymmetric Yang-Mills theory in $d=p+1$ dimensions has 16 real supercharges and consists of a vector multiplet transforming in the adjoint representation of the gauge group $G$. The fields in this multiplet are the gauge field $A_\mu$, $9-p$ real scalar fields, $\Phi_m$, and 16 fermionic degrees of freedom, or gaugini, collectively denoted by $\Psi$. Depending on the  dimension and signature of space-time the fermionic degrees of freedom are arranged into spacetime spinors as summarized in Table \ref{tabspinors}.
\begin{table}\centering
\begin{tabular}{c|ll}
Dimensions & Lorentzian & Euclidean \\
\hline
7& 1 & 2 (Majorana)\\
6& 2\p{1} (Weyl)& 2 (Majorana)\\
5& 2 & 2\\
4& 4\p{1} (Majorana)  & 4 (Weyl)\\
3& 8\p{1} (Majorana) & 4\\
2 & 16 (Majorana-Weyl) & 8 (Majorana)
\end{tabular}
\caption{\label{tabspinors}Number of minimal spinors in each dimension used in Lorentzian and Euclidean field theories. The conditions the spinors satisfy are indicated in brackets. In all cases we denote the collective 16-component fermion with the symbol $\Psi$.}
\end{table}
The index $m$, which labels the scalar fields, transforms in the fundamental representation of the R-symmetry group, which is $\SO(9-p)$ for Lorentzian\footnote{We work with a ``mostly +'' signature.} theories and $\SO(1,8-p)$ for the Euclidean ones. We are mostly interested in  Euclidean theories, as we intend to study SYM on $S^{p+1}$, but for the moment we keep the discussion general and discuss both cases. 

The classical action for the $(p+1)$-dimensional maximal SYM theory on flat space can be derived by dimensional reduction of the unique SYM action in ten dimensions, see for instance \cite{Brink:1976bc}. Note that to obtain the Euclidean theory one must perform a timelike dimensional reduction. Explicitly the Lagrangian of the $d$-dimensional SYM theory on flat space reads\footnote{Since we are interested in QFTs we take $p\geq 1$. We also set the $\theta$-term in the four-dimensional SYM action to zero.}
\bea\label{LSYMflat}
{\cal L}_\text{SYM}&=&\f1{2g_{{\rm YM}}^2}\Tr\left[- F_{\mu\nu}F^{\mu\nu} -D_\mu \Phi_m D^\mu \Phi^m + \bar{\Psi} \gamma^\mu D_\mu\Psi\right.\nonumber\\
&&\hspace{3cm}\left.- \f12[\Phi_m,\Phi_n][\Phi^m,\Phi^n]  +\bar{\Psi}\Gamma^m[\Phi_m,\Psi]\right]~.
\eea
Here $\bar{\Psi}=\Psi^\dagger \Gamma_0$ is the Dirac adjoint with $\Gamma_0$ the ten-dimensional gamma matrix along the time direction.\footnote{We use conventions in which the $(9-p)$-dimensional gamma matrices are denoted by $\Gamma_m$ and the $(p+1)$-dimensional ones are $\gamma_{\mu}$. The Clifford algebra is $\{\gamma_{\mu},\gamma_{\nu}\} = 2g_{\mu\nu}$, with $g_{\mu\nu}$ the metric on space-time.} The Yang-Mills field strength is given by $F_{\mu\nu} = 2\partial_{[\mu}A_{\nu]}+[A_\mu,A_\nu]$, and the gauge covariant derivatives are
\be
D_\mu\Phi_m = \partial_\mu \Phi_m + [A_\mu,\Phi_m]~,\qquad D_\mu \Psi = \partial_\mu \Psi +[A_\mu,\Psi]~.
\ee
The $m,n$ indices are raised and lowered with the flat metric on ${\bf R}^{9-p}$ for Lorentzian theories or ${\bf R}^{1,8-p}$ for Euclidean theories.  For Euclidean theories this implies that one of the scalar fields, $\Phi^0$, has the wrong sign kinetic term. Notice also that the scalar-fermion interaction term involves internal gamma matrices, $\Gamma^m$, associated with the R-symmetry.

The Yang-Mills coupling constant, $g_{{\rm YM}}$, is dimensionful for $d=p+1\ne 4$, and the mass dimension is given by $[g_\text{YM}^{2}] = 3-p$. This means that maximal SYM theories are non-renormalizable for $p>3$ and have to be incorporated in a UV complete theory at high energies. Indeed, maximal SYM in five dimensions, i.e. $p=4$, is conjectured to grow an extra dimension at high energies and flow towards the (2,0) CFT in six dimensions \cite{Douglas:2010iu,Lambert:2010iw}. Maximal SYM in six dimensions is expected to be UV completed by a non-gravitational but non-local theory called little string theory. Little string theory comes in two flavors, depending on the chirality of the supercharges in six dimensions. Six-dimensional SYM has $(1,1)$ supersymmetry and therefore flows towards the corresponding $(1,1)$ little string theory in the UV, see \cite{Aharony:1999ks,Kutasov:2001uf} for reviews and further references. For $p>5$ maximal SYM has a UV completion within string theory as the worldvolume theory on D$p$-branes. For $p\leq3$ the UV physics is under better control. When $p=3$ it is well-known that maximal SYM is conformally invariant and thus UV finite. For $p<3$ the YM coupling is asymptotically free and the physics at low-energies is strongly coupled. The theory in three dimensions, i.e. $p=2$, is believed to flow to the interacting ABJM CFT which has maximal supersymmetry and describes the low-energy dynamics of M2-branes \cite{Aharony:2008ug}. The IR dynamics of the maximal two-dimensional SYM is somewhat more involved, see \cite{Kologlu:2016aev} for a recent discussion.

Placing SYM on curved backgrounds such as $S^{p+1}$ via a minimal coupling, i.e. replacing $\eta_{\mu\nu}$ with $g_{\mu\nu}$ and partial with covariant derivatives in \eqref{LSYMflat}, results in an action that in general does not posses any supersymmetry. This is because constant supersymmetry transformation parameters, $\epsilon$, do not exist on a general curved manifold. The supersymmetry transformation of the action is proportional to the derivative of $\epsilon$ which in general does not vanish. Understanding which supersymmetric QFTs can be placed on which curved manifolds while preserving some amount of supersymmetry can be done systematically using the formalism described in \cite{Festuccia:2011ws}. For maximal SYM on $S^{p+1}$ this question was addressed in the earlier work \cite{Blau:2000xg}, see also \cite{Pestun:2007rz,Minahan:2015jta}.

In this paper we are interested in the maximal SYM theory placed on $S^{p+1}$ with metric $\mathcal{R}^2\dd \Omega_{p+1}^2$ where $\mathcal{R}$ is the radius of the sphere and $\dd \Omega_{p+1}^2$ is the unit radius\footnote{The Ricci scalar of the sphere with metric $\mathcal{R}^2\dd \Omega_{p+1}^2$ is equal to $\frac{p(p+1)}{\mathcal{R}^2}$.} Einstein metric on $S^{p+1}$. It was shown in \cite{Blau:2000xg} that these Euclidean theories can preserve 16 real supercharges and the supersymmetry parameter obeys the equation
\be\label{spinorderivative}
\nabla_\mu \epsilon =\frac{1}{2\mathcal{R}} \gamma_\mu \Gamma \epsilon\quad \text{with}\quad \Gamma = \Gamma^{012}\,,
\ee
This construction only works for $p\le 6$. This is closely related to the fact that superconformal algebras only exist in six or fewer dimensions. The action for maximal SYM on $S^{p+1}$ is explicitly given as a deformation of the action in \eqref{LSYMflat}. First we have to introduce a minimal coupling of the Lagrangian in \eqref{LSYMflat} to the metric on the sphere. In addition, to ensure that the the supersymmetry generated by the spinor in \eqref{spinorderivative} is preserved, we have to add the following extra terms to the Lagrangian
\bea
\delta {\cal L} &=&  -\frac{1}{\mathcal{R}^2} \Tr \left[(p-1)\Phi_m\Phi^m +(p-3)\Phi_a\Phi^a\right]\nonumber\\
&&\qquad+ \frac{1}{2\mathcal{R}}(p-3)\Tr\left[\bar{\Psi}\Gamma\Psi - 8\Phi_0[\Phi_1,\Phi_2]\right]~,\label{symlagrangian}
\eea
where the index $a$ only runs from $0$ to $2$ and is contracted using the Lorentzian metric just like the $m,n$ indices \cite{Blau:2000xg}. The Lagrangian in \eqref{symlagrangian} for $p\ne 3,6$ breaks the $\SO(1,8-p)$ R-symmetry group of the maximal theory in Euclidean space to the subgroup\footnote{For $p=6$ the R-symmetry group is $\SO(1,2)\simeq \SU(1,1)$ and is preserved by the Lagrangian in \eqref{symlagrangian}.}
\be\label{NonCompactR}
\SU(1,1)\times \SO(6-p)~.
\ee
This symmetry breaking pattern is an important guiding principle for constructing the spherical brane solutions of ten-dimensional supergravity.

There are at least three important reasons to study SYM theories on $S^{p+1}$. First, this is a maximally symmetric space which is also the unique curved manifold on which one can preserve 16 supercharges. Placing a supersymmetric theory on a sphere is an essential ingredient in the context of supersymmetric localization and indeed it was recently shown in \cite{Minahan:2015jta}, following the seminal work \cite{Pestun:2007rz}, how to study the partition function of maximal SYM using this method for any $1\leq p\leq 6$. This in turn paves the way to compute exactly certain supersymmetric correlation functions of the SYM theory. Finally, the radius, $\mathcal{R}$, of the sphere provides a natural IR cut-off for the dynamics of the SYM theory which is compatible with supersymmetry. This is especially important in the holographic context where the IR physics of SYM theories for $p\neq 3$ results in singularities of the dual supergravity solutions. These can be resolved by introducing finite temperature in the form of a black hole horizon \cite{Itzhaki:1998dd}. The finite temperature is a convenient IR regulator which however breaks supersymmetry completely. As we show below, our spherical D$p$-brane solutions, which are holographically dual to the maximal SYM on $S^{p+1}$, are regular in the IR while preserving all 16 supercharges.

\section{D-branes in flat space}
\label{sec:flatDp}

Before embarking on the journey towards constructing supergravity solutions describing D$p$-branes with a spherical worldvolume we first review the physics of their flat counterparts. Black branes are solution of ten-dimensional type II supergravity that source the metric, the dilaton, as well as $(p+2)$-form field strengths \cite{Horowitz:1991cd}. The solutions are characterized by their conserved electric charge $\mu_p$ and the ADM tension $T_p$. These branes break either all of the original 32 supercharges of type II supergravity or only half of them. We will focus on the latter case for which the branes are extremal in the sense that their tension equals their charge, i.e. $T_p=\mu_p$. Within string theory these supergravity backgrounds are interpreted as the backreaction of a large number of fundamental D$p$-branes on the ten-dimensional geometry in which they are immersed \cite{Polchinski:1995mt}. This interpretation has passed many consistency checks in string theory and ultimately led to the AdS/CFT correspondence. The low-energy physics on the flat worldvolume of the fundamental D$p$-branes is described by maximal SYM in flat space. This implies that upon gravitational backreaction the supergravity solution describing supersymmetric black branes is holographically dual to this SYM theory as first suggested in \cite{Itzhaki:1998dd}.

The full ten-dimensional supergravity solution describing D$p$-branes with flat worldvolume, with $p\le 6$, in asymptotically flat space and in string frame is (see for example \cite{Blumenhagen:2013fgp}) 
\bea
\dd s_{10}^2 &=& H^{-1/2}\dd s_{p+1}^2 + H^{1/2}\dd s_{9-p}^2\,,\label{eq:flatDpbranes}\\
\e^\Phi &=& g_s H^{(3-p)/4}\,,\\
C_{p+1} &=& (g_s H)^{-1}~\vol_{p+1}\,.
\eea
Here $\dd s_{p+1}^2$ and $\dd s_{9-p}^2$ denote the flat metrics on ${\bf R}^{1,p}$ and ${\bf R}^{9-p}$ respectively, $\vol_{p+1}$ is the volume form on ${\bf R}^{1,p}$ and $H$ is a harmonic function on ${\bf R}^{9-p}$. The harmonic function has isolated singularities at the position of the branes. For a single stack of D$p$-branes at the origin we have $\dd s_{9-p}^2 = \dd r^2 + r^2 \dd \Omega_{8-p}^2$, with $ \dd \Omega_{8-p}^2$ the unit radius metric on a round $8-p$ sphere. The harmonic function in this case is
\be\label{flatharmonic}
H= 1 + \f{g_s N}{\mu_{6-p}V_{6-p}r^{7-p}}\,,
\ee
where $V_{n-1} = 2\pi^{n/2}/\Gamma(n/2)$ is the volume of the unit radius $n$-sphere. The fundamental charge of a D$p$-brane is given by\footnote{The ten-dimensional Newton constant is related to the string length $\ell_s$ through $4\pi\kappa_{10}^2 = (2\pi \ell_s)^8$, therefore $2\kappa_{10}^2 \mu_p\mu_{6-p} = 2\pi$. Note that in our conventions the Newton constant does not depend on $g_s$.}
\be\label{eq:mupdef}
\mu_p =\f{2\pi}{(2\pi\ell_s)^{p+1}}\,,
\ee
and the Yang-Mills coupling constant of the worldvolume gauge theory is
\be
g_{\rm YM}^2 =\f{(2\pi)^2g_s}{(2\pi\ell_s)^4\mu_p}~.
\ee
The constants in \eqref{flatharmonic} must satisfy a Dirac quantization condition. Indeed, integrating the magnetic field strength over $\dd \Omega_{8-p}^2$ leads to
\be
\f{1}{2\kappa_{10}^2\mu_p}\int \star \dd C_{p+1} = \f{N(7-p)V_{8-p}}{2\kappa_{10}^2\mu_p\mu_{6-p}V_{6-p}} = N\,,
\ee
where the integer $N$ is interpreted as the number of D$p$-branes.

The field theory limit of D$p$-branes in a holographic context was first studied in \cite{Itzhaki:1998dd} (see also \cite{Boonstra:1998mp}). Introducing the dimensionless radial coordinate $U=r/(2\pi \ell_s)$, this limit is equivalent to zooming in on the near-horizon region of the branes: 
\be\label{eq:nhlimit}
\f{g_s N ~ U^{p-7}}{2\pi V_{6-p}}   \gg 1\,.
\ee
Using \eqref{eq:nhlimit}, the metric and dilaton simplify to
\bea
\dd s_{10}^2 &=& (g U)^{\f{7-p}{2}}~\dd s_{p+1}^2 + (g U)^{\f{p-7}{2}}\left(\dd U^2 +U^2 \dd \Omega_{8-p}^2\right)\,,\label{nearhorizonbranesmetric}\\
\e^\Phi &=& g_s (g U)^{\f{(p-3)(7-p)}{4}} \,,\label{nearhorizonbranesdilaton}\\
C_{p+1} &=& g_s^{-1} (g U)^{7-p} \vol_{p+1}\,,\label{nearhorizonbranesC}
\eea
where we have introduced $g$ as\footnote{The real constant $g$ will be identified with the coupling constant of the $(p+2)$-dimensional gauged supergravity theory in which the brane solutions can be effectively described.}
\begin{equation}\label{gsugra}
(2\pi \ell_s g)^{p-7} = \f{g_s N}{2\pi V_{6-p}}\,.
\end{equation}

For $0<p<3$ at high energies, $ U\gg 1$, the string coupling becomes small indicating that the theory is free in the UV. As discussed in Section~\ref{sec:fieldtheory} this is the expected UV behavior of maximal SYM theory in $1<d<4$ dimensions. Conversely for $3<p<7$ the dilaton increases at high energies indicating that the field theory is strongly coupled. This again fits nicely with the fact that for $d>4$ the SYM theory is not renormalizable. Clearly the case $p=3$ is special since the string coupling is constant throughout the solution and the metric is that of AdS$_5\times S^5$. This is the well-known holographic dual description of the conformal $\mathcal{N}=4$ SYM theory in $d=4$. The background in \eqref{nearhorizonbranesmetric}-\eqref{nearhorizonbranesC}  possesses $\ISO(1,p)\times\SO(9-p)$ isometry for $p\neq 3$ and $\SO(2,4)\times \SO(6)$ for $p=3$. This is the same as the global symmetry group of the SYM theories discussed in Section~\ref{sec:fieldtheory}. It is therefore clear that this near-horizon solution nicely exhibits the physics we expect from a holographic dual to SYM on flat space. We refer to \cite{Itzhaki:1998dd} and references thereof for further support of this holographic duality.

Our goal is to generalize the solutions in \eqref{nearhorizonbranesmetric}-\eqref{nearhorizonbranesC} and construct supergravity backgrounds which correspond to spherical D$p$-branes and provide a holographic description of maximal SYM on $S^{p+1}$. This necessitates an understanding of how to construct supergravity solutions for D-branes with Euclidean worldvolume. This was addressed in several papers by Hull \cite{Hull:1998vg,Hull:1998fh,Hull:1998ym} where he argued that there are Euclidean branes, or E-branes, not of regular type II string theory but of the so-called type II$^*$ string theory. The existence of a low-energy supergravity limit of these type II$^*$ string theories can be deduced independently from a supergravity point of view \cite{Bergshoeff:2000qu}. The type II$^*$ supergravity theories admit E-brane solutions\footnote{Note that in the notation of \cite{Hull:1998vg,Hull:1998fh,Hull:1998ym} an E$(p+1)$-brane is the Euclidean version of a D$p$-brane.} for which the brane worldvolume is Euclidean and the time direction is transverse to the brane worldvolume, i.e. E-branes resemble instantons. The E-brane solutions can be obtained from the D$p$-brane solutions above by analytically continuing the time direction of the brane worldvolume into a spatial coordinate and at the same time analytically continuing the polar angle of the sphere transverse to the brane into a time-like coordinate. This analytic continuation results in changing the worldvolume of the brane from $\mathbf{R}^{1,p}$ to $\mathbf{R}^{p+1}$ and the transverse $S^{8-p}$ sphere in \eqref{nearhorizonbranesmetric} to de Sitter space, dS$_{8-p}$. The analytic continuation does not only affect the metric, but also changes the R-R fields. In \cite{Hull:1998vg,Hull:1998fh,Hull:1998ym} all R-R fields are taken to be real with ``wrong sign'' kinetic terms. In this paper we use an equivalent formulation in which all R-R fields are imaginary with ``usual sign'' kinetic terms. To ensure supersymmetry the Killing spinors for these E-branes have to satisfy rather unusual reality conditions, this is explained in some detail in Appendix \ref{App:flatEbranes}. Finally we note that solutions of the Lorentzian type IIA$^*$ string theory should uplift to solutions of the so-called M$^*$ theory, see \cite{Hull:1998vg,Hull:1998fh,Hull:1998ym}, which has the somewhat exotic $(2,9)$ signature of the metric, i.e. two time-like and nine spatial dimensions.

\section{Spherical branes}
\label{sec:SpherDp}

To construct the near-horizon solution for Euclidean D$p$-branes wrapped on spheres we can utilize intuition from the field theory discussion in Section \ref{sec:fieldtheory} and make a suitable ansatz for the ten-dimensional metric. The total isometry group of the solution should be a direct product of the isometry group of the $S^{p+1}$ which the D$p$-brane is wrapping and the R-symmetry group of the Yang-Mills field theory in $p+1$ dimensions:
\be\label{eq:isomgroup}
\SO(p+2)\times \SU(1,1)\times \SO(6-p)\,.
\ee
A ten-dimensional metric ansatz that implements these symmetries is given by
\be\label{metricansatz10d}
\dd s_{10}^2 = \Delta\left[\dd r^2 + {\cal R}^2\e^{2A} \dd \Omega_{p+1}^2 + \e^{2B}\left(\dd \theta^2 + P \cos^2\theta\,\dd \widetilde{\Omega}_2^2 + Q\sin^2\theta\,\dd \Omega_{5-p}^2\right)\right]\,.
\ee
The functions $A$, $B$, $\Delta$, $P$, and $Q$ depend on $r$ and $\theta$ and satisfy suitable positivity conditions such that the metric has the correct signature, while $\mathcal{R}$ is a constant that sets the radius of $S^{p+1}$. The metric on a round $n$-sphere is denoted by $\dd \Omega_n^2$ with volume form $\vol_n$. Clearly the $\dd \Omega_{p+1}^2$ and $\dd \Omega_{5-p}^2$ factors in the metric realize the $\SO(p+2)\times \SO(6-p)$ part of the isometry group in \eqref{eq:isomgroup}. The non-compact $\SU(1,1)$ factor in the R-symmetry of the SYM theory is realized as the isometry group of two-dimensional de Sitter space with metric
\be\label{eq:metdS2}
\dd \widetilde{\Omega}_2^2 = -\dd t^2 + \cosh^2 t\, \dd \psi^2\,,
\ee
where $\psi$ is $2\pi$-periodic. 

Note that for $P=Q=1$ the metric in \eqref{metricansatz10d} simplifies significantly, namely the metric transverse to the worldvolume of the branes is the round metric on dS$_{8-p}$ which has $\SO(1,8-p)$ isometry group, i.e. the same as for the Euclidean D$p$-branes in flat space discussed at the end of Section~\ref{sec:flatDp}. Even without having an explicit solution for spherical branes, intuition from field theory suggests that for values of the radial coordinate much larger than the scale set by $\mathcal{R}$ the supergravity solution should reduce precisely to the Euclidean D$p$-brane background with $P=Q=1$. This is suggested by the UV limit in the field theory where the curvature of $S^{p+1}$ should play no role in the dynamics of the SYM theory and it should reduce to that in flat space.
 
In addition to the metric \eqref{metricansatz10d}, we also have to make an ansatz for the type II NS-NS and R-R form fields and the dilaton. D$p$-branes are electrically charged under $C_{p+1}$ and it is natural to expect there to be a component of $C_{p+1}$ along the $\dd \Omega_{p+1}^2$ worldvolume of the branes. We should also allow for all other form fields in the supergravity to have non-zero values as long as they preserve the isometry group in \eqref{eq:isomgroup}. In addition the dilaton can be an arbitrary function of $r$ and $\theta$. With this ansatz at hand one should analyze carefully the equations of motion and the supersymmetry variations of the ten-dimensional supergravity theory imposing that the background preserves 16 out of the 32 supercharges. This analysis should result in a system of coupled non-linear partial differential equations for the unknown functions in the ansatz. It is fair to assume that without any further insight it will be difficult to solve explicitly this system of equations. Fortunately progress can be made by employing a well-tested strategy in top-down holography, namely reduce the ten-dimensional problem to a supergravity problem in $p+2$ dimensions. This can be achieved by employing a consistent truncation of the ten-dimensional supergravity to an appropriate gauged supergravity in $p+2$ dimensions.

\subsection{Supergravity in $p+2$ dimensions}
\label{subsec:Sugrap+2}

The gauged supergravity theories of interest are maximally supersymmetric and arise as consistent truncations of type II supergravity on $S^{8-p}$. The vacua of these supergravities are directly related to the field theory limits of D$p$-branes discussed in Section \ref{sec:flatDp} and thus for $p\ne3$ the vacuum breaks half of the supersymmetries. In order to describe spherical branes we must analytically continue these gauged supergravity theories so as to work with an Euclidean theory. After constructing the solutions of interest we can uplift them to ten dimensions where we recover the time direction, as in \eqref{metricansatz10d}, and thus obtain a fully Lorentzian solution of type II supergravity.  In this section we start by briefly describing the Lorentzian gauged supergravity theories before performing the analytic continuation. Since the construction of the spherical brane solutions proceeds similarly in different dimensions we present a uniform description of the Lagrangian and BPS equations for all values of $p$. In Appendix~\ref{App:sugra} we give a more detailed discussion of the supergravity theories in various dimensions used in this paper and their analytic continuation. 

The field theory discussion in Section~\ref{sec:fieldtheory} suggest that to construct the spherical brane solutions of interest we can restrict to an $\SU(1,1)\times \SO(6-p)$ invariant truncation of the maximally supersymmetric gauged supergravity theory. This ensures that the R-symmetry of the SYM theory, realized as a gauge symmetry in the supergravity theory, is preserved. In addition we are interested in supergravity solutions which preserve the $\SO(p+2)$ isometry of the sphere which the branes are wrapping. This in turn implies that all  fields present in the gauged supergravity theory except the metric and scalar fields should be set to zero. As discussed in detail in Appendix~\ref{App:sugra}, imposing these symmetries on the gauged supergravity leads to a consistent truncation which includes only the metric and three real scalar fields: the ``dilaton'' $\phi$, a real scalar $x$ and a pseudoscalar $\chi$.\footnote{The cases $p=3$ and $p=6$ are somewhat special and will be discussed separately below.} These scalar fields have a nice interpretation in the SYM theory on $S^{p+1}$. The dilaton is dual to the Yang-Mills coupling, the scalar field $x$ is dual to the bosonic bilinear mass term $\Phi^2$, and the pseudoscalar is dual to the fermionic bilinear mass term $\Psi^2$ which appear in the field theory Lagrangian \eqref{symlagrangian}. It turns out that it is more convenient to work with the scalar fields $\lambda$ and $\beta$ (discussed further in Appendix~\ref{App:sugra}) which are linear combinations of the scalar fields $x$ and $\phi$. In terms of these fields, the bosonic actions for the truncated gauged supergravity theories take the following uniform form for $0<p<6$
\be\label{sugraaction}
S = \f{1}{2\kappa_{p+2}^2}\int\star_{p+2}\left\{ R + \f{3p}{2(p-6)}|\dd \lambda|^2 - \f{1}{2}\left(|\dd\beta|^2 + \e^{2\beta}|\dd \chi|^2\right)- V\right\}\,,
\ee
where $V$ is the scalar potential. It is clear from the kinetic terms that $(\beta,\chi)$ span an SL(2)$/$SO(2) coset which can be conveniently parametrized by a single complex scalar 
\be\label{taudef}
\tau \equiv \chi+i\e^{-\beta}\,.
\ee
The kinetic term for $\tau$ can then be written in terms of the K{\"a}hler potential 
\be\label{eq:Kpotdef}
{\cal K} = -\log\left(\frac{\tau-\bar{\tau}}{2}\right) = \beta\,,
\ee
as
\be\label{eq:Kahlerkin}
{\cal K}_{\tau\bar{\tau}}|\dd \tau|^2 = \f{1}{4}\left(|\dd\beta|^2 + \e^{2\beta}|\dd \chi|^2\right)\,,
\ee
where ${\cal K}_{\tau\bar{\tau}} = \partial_\tau\partial_{\bar \tau} {\cal K}$ is the K{\" a}hler metric. The scalar potential can be compactly expressed in terms of a superpotential which is holomorphic in $\tau$ and reads:
\be\label{superpotential}
{\cal W} = \left\{\begin{array}{ll} 
-g~\e^{\f{1}{2}\lambda}\left(3\tau + (6-p)i\e^{-\f{p}{6-p}\lambda}\right) & \text{for $p<3$\,,}\\
-g~\e^{\f{3(2-p)}{2(6-p)}\lambda}\left(3i\e^{\f{p}{6-p}\lambda} + (6-p)\tau\right) & \text{for $p>3$\,.}\end{array}\right.
\ee
Here $g$ is the $\SO(9-p)$ gauge coupling constant of the maximal gauged supergravity theory. The scalar potential is given by
\be\label{eq:defVgenp}
V = \f12 \e^{\cal K}\left(\f{6-p}{3p}\left|\partial_\lambda {\cal W}\right|^2 + \f14{\cal K}^{\tau\bar{\tau}}D_\tau {\cal W}D_{\bar \tau} \overline{\cal W} - \f{p+1}{2p}\left| {\cal W}\right|^2\right)\,,
\ee
where $D_a = \partial_a + \partial_a {\cal K}$ is the K{\"a}hler covariant derivative. 

It is clear from the kinetic term of the scalar $\lambda$ in \eqref{sugraaction} as well as the superpotential in \eqref{superpotential} that $p=6$ has to be treated separately. This is in harmony with the field theory discussion in Section~\ref{sec:fieldtheory} where it was shown that the R-symmetry is unbroken upon placing SYM on $S^7$. This in turn implies that in the supergravity theory we should retain only the complex scalar field, $\tau$, and not include the scalar $\lambda$. The eight-dimensional gravitational action for $p=6$ then reads\footnote{Again we refer to Appendix~\ref{App:sugra} for more details on how to obtain this action from maximal supergravity in eight dimensions.}
\be\label{eq:8dsugra}
S = \f{1}{2\kappa_{8}^2}\int\star_8\left\{ R - 2{\cal K}_{\tau\bar{\tau}}|\dd \tau|^2- V\right\}\,,
\ee
where
\be\label{eq:defVp=6}
V = \f12 \e^{\cal K}\left(\f14{\cal K}^{\tau\bar{\tau}}D_\tau {\cal W}D_{\bar \tau} \overline{\cal W} - \f{7}{12}\left| {\cal W}\right|^2\right)\,,
\ee
and
\be\label{eq:Wp=6}
{\cal W} = -3ig\,.
\ee
It is reassuring to observe that the supergravity action in \eqref{eq:8dsugra} can be obtained by a formal limit of the action in \eqref{sugraaction} by taking $\lambda/(p-6)\to 0$ and $p\to6$.

As mentioned above, due to the runaway potential for the dilaton $\phi$ there is no vacuum solution of the gravitational theories in \eqref{sugraaction} and \eqref{eq:8dsugra} that preserves all 32 supersymmetries.\footnote{The case $p=3$ is an exception.} There are however domain wall solutions which preserve 16 supercharges and are closely related to the flat brane solutions in ten dimensions discussed in Section~\ref{sec:flatDp}, see for example \cite{Boonstra:1998mp}. These solutions are obtained by setting $\chi=x=0$, or equivalently $\beta  = \f{p}{p-6}\lambda$, and read
\be\label{flatdomainwall}
\dd s^2_{p+2} = \dd r^2 + \e^{\f{2(9-p)}{(6-p)(p-3)}\lambda}\dd s_{p+1}^2\,,\quad 
\e^{\f{(p-3)}{6-p}\lambda} = \f{g(3-p)^2}{2p}(r-r_0)\,,
\ee
where $\dd s_{p+1}^2$ is the flat metric on Minkowski space and $r_0$ is an integration constant that can be set to zero by shifting appropriately the radial coordinate $r$. These solutions can be uplifted to solutions of type II supergravity using the uplift formulae discussed in Section~\ref{subsec:uplift}. The end result of this uplift is given by the following ten-dimensional background
\bea
\dd s^2 &=& \e^{\lambda}\left(\dd s^2_{p+2} +\f{1}{g^2}\e^{\f{2(p-3)}{6-p}\lambda}\dd \Omega_{8-p}^2\right)\,,\\
\e^\Phi &=& g_s\e^{\f{p(7-p)}{2(6-p)}\lambda}\,,\\
F_{8-p} &=& \f{7-p}{g_sg^{7-p}}\vol_{8-p}\,.
\eea
This solution precisely matches the near-horizon limit of the flat D$p$-brane solutions in \eqref{nearhorizonbranesmetric}-\eqref{nearhorizonbranesdilaton} where $gU = \e^{\f{2p}{(6-p)(p-3)}\lambda}$ and the number of branes $N$ is related to the supergravity coupling constant $g$ via \eqref{gsugra}.

So far we have discussed only Lorentzian supergravities. However the spherical branes of interest here have a Euclidean worldvolume and thus should be described by Euclidean gauged supergravity theories. Such theories should be maximally supersymmetric with an $\SO(1,8-p)$ gauge group and should be closely related to the more familiar $\SO(9-p)$ maximal gauged supergravity theories in Lorentzian signature. These Euclidean supergravity theories are unfortunately not available in the literature. We resolve this impasse by performing an analytic continuation of the truncated Lorentzian supergravity theories described by the Lagrangians in \eqref{sugraaction} and \eqref{eq:8dsugra}.

At the level of the action the analytic continuation is straightforward. The metric becomes Euclidean and the only real modification to the action stems from the fact that the pseudo scalar $\chi$ becomes imaginary
\be\label{eq:chianalytic}
\chi \to i\chi\,.
\ee
This results in the ``wrong sign'' kinetic term for $\chi$. The scalar $\tau$ in \eqref{taudef} appears to be a purely imaginary scalar field and thus is not appropriate to describe two independent scalar fields. We must therefore consider $\tau=i( \chi+\e^{-\beta})$ and, what used to be, its complex conjugate $\tilde{\tau}=i( \chi-\e^{-\beta})$ as two independent scalar fields in the Euclidean theory.\footnote{This is a familiar predicament from similar constructions of Euclidean supergravity solutions in a holographic context \cite{Freedman:2013ryh,Bobev:2013cja}.} Similarly we should work with \emph{two} superpotentials, ${\cal W}$ as defined in \eqref{superpotential} and $\widetilde{\cal W}$ obtained by complex conjugation of ${\cal W}$ accompanied by the replacement $\bar\tau \to \tilde{\tau}$,
\be\label{tildesuperpotential}
\widetilde{\cal W} = \left\{\begin{array}{ll} -g~\e^{\f{1}{2}\lambda}\left(3\tilde\tau - (6-p)i\e^{-\f{p}{6-p}\lambda}\right) & \text{for $p<3$\,,}\\
g~\e^{\f{3(2-p)}{2(6-p)}\lambda}\left(3i\e^{\f{p}{6-p}\lambda} - (6-p)\tilde\tau\right) & \text{for $p>3$\,.}\end{array}\right.
\ee
The scalar potential of the Euclidean theory is obtained by replacing $\overline{\cal W}$ by $\widetilde{\cal W}$ in \eqref{eq:defVgenp}. 

With this Euclidean supergravity theory at hand we are now in a position to discuss how to construct the spherical branes of interest. We start by writing the following metric ansatz compatible with the spherical symmetry of the worldvolume of the brane
\be\label{eq:genpmetAnsatz}
\dd s_{p+2}^2 = \dd r^2 + {\cal R}^2 \e^{2A} \dd \Omega_{p+1}^2\,.
\ee
In addition we assume that the scalar fields and the warp factor $A$ only depend on the radial variable $r$. The constant ${\cal R}$ should be thought of as the radius of $S^{p+1}$ and is auxiliary since it can be absorbed into a redefinition of metric function $A$.\footnote{To stay in the regime of validity of supergravity we have to make sure that $\mathcal{R} \e^{A}$ is larger than the Planck and the string scales throughout the solution.} To obtain the brane solutions with flat worldvolume one should take ${\cal R}\to\infty$. As we shall discuss below this is not a smooth limit, nevertheless it still proves useful to keep the constant ${\cal R}$ explicitly in the formulae below.

Equipped with this ansatz we can plug it in the supersymmetry variations of the $(p+2)$-dimensional gauged supergravity theory and look for solutions which preserve 16 supercharges. This is discussed in some detail in Appendix~\ref{App:sugra}. The end result is the following system of BPS equations which should be obeyed by the metric function and the three scalar fields:
\bea
(\lambda')^2 &=& \e^{{\cal K}}\left(\f{6-p}{3p}\right)^2(\partial_\lambda{\cal W})(\partial_\lambda\widetilde{\cal W})\,,\label{lambdaeq}\\
(\lambda')(\tau')&=&\e^{{\cal K}}\left(\f{6-p}{12p}\right)\left(\partial_\lambda {\cal W}\right){\cal K}^{\tau\tilde{\tau}}D_{\tilde \tau} \widetilde{\cal W} \,,\label{taueq}\\
(\lambda')(\tilde\tau')&=&\e^{{\cal K}}\left(\f{6-p}{12p}\right)(\partial_\lambda \widetilde{\cal W}){\cal K}^{\tilde{\tau}\tau}D_{\tau} {\cal W} \,,\label{tautildeeq}
\eea
\bea
(\lambda')(A' - {\cal R}^{-1}\e^{-A})&=& -\e^{{\cal K}}\left(\f{6-p}{6p^2}\right)(\partial_\lambda {\cal W})\widetilde{\cal W}\,,\label{Aeq1}\\
(\lambda')(A' + {\cal R}^{-1}\e^{-A})&=& -\e^{{\cal K}}\left(\f{6-p}{6p^2}\right)(\partial_\lambda \widetilde{\cal W}){\cal W}\,,\label{Aeq2}
\eea
where ${\cal K}^{\tilde{\tau}\tau}$ is the inverse of the K\"ahler metric in \eqref{eq:Kahlerkin}. Equations \eqref{lambdaeq}, \eqref{taueq}, and \eqref{tautildeeq} arise from the spin-$\frac{1}{2}$ supersymmetry variations of the gauged supergravity theory, while \eqref{Aeq1} and \eqref{Aeq2} arise from the spin-$\frac{3}{2}$ variations.

Equations \eqref{Aeq1} and \eqref{Aeq2} lead to a first order differential equation together with the following algebraic relation for the metric function $A(r)$
\begin{equation}\label{eq:algebraicA}
 \e^{A} = \frac{1}{{\cal R}g^2}\frac{2p}{6-p} \frac{\tilde\tau-\tau}{\tilde\tau+\tau}\e^{\frac{2(p-3)}{6-p}\lambda}(\lambda')\,.
\end{equation}
Fortunately these two equations are compatible with each other. In addition one can explicitly check that all BPS equations in \eqref{lambdaeq}-\eqref{Aeq2} are compatible with the second order equations of motion derived from the action in \eqref{sugraaction} after the analytic continuation in \eqref{eq:chianalytic}.\footnote{The case $p=6$ should again be treated separately and is discussed in more detail in Appendix~\ref{appB:D6}.}

Note that upon taking the limit ${\cal R}\to \infty$ in \eqref{lambdaeq}-\eqref{Aeq2} accompanied with $\tau=\tilde\tau$, which in turn implies ${\cal W}=\widetilde{\cal W}$, we obtain the BPS equations for a domain wall with flat slices. These equations are then solved by the Euclidean analog of \eqref{flatdomainwall}.

\subsection{Analysis of the BPS equations}

We now perform a preliminary analysis of the BPS equations \eqref{lambdaeq}-\eqref{Aeq2}. It proves convenient to introduce a new parametrization of the scalar fields given by
\be\label{tautoXY}
\begin{array}{ll}
\tau = i\e^{-\f{p}{6-p}\lambda}(X+Y)\,,\quad \tilde\tau = -i\e^{-\f{p}{6-p}\lambda}(X-Y)\,,&\text{for $p<3$}\,,\\
\tau = i\e^{\f{p}{6-p}\lambda}(X+Y)\,,\quad \tilde\tau = -i\e^{\f{p}{6-p}\lambda}(X-Y)\,,&\text{for $p>3$}\,.
\end{array}
\ee
When the BPS equations are solved it is important to impose appropriate boundary conditions in the IR. The physics of the SYM theory on $S^{p+1}$ suggests that the supergravity solutions should cap off smoothly in the IR and it is thus natural to look for solutions in which close to some finite value of the radial coordinate $r \to r_{{\rm IR}}$ the metric looks like the metric on $(p+2)$-dimensional flat space in spherical coordinates
\be
\dd s_{p+2}^2 \approx \dd r^2 + (r-r_\text{IR})^2 \dd \Omega_{p+1}^2\,.
\ee
%
\begin{figure}[h]
\centering
\begin{overpic}[width=0.4\textwidth,tics=10]{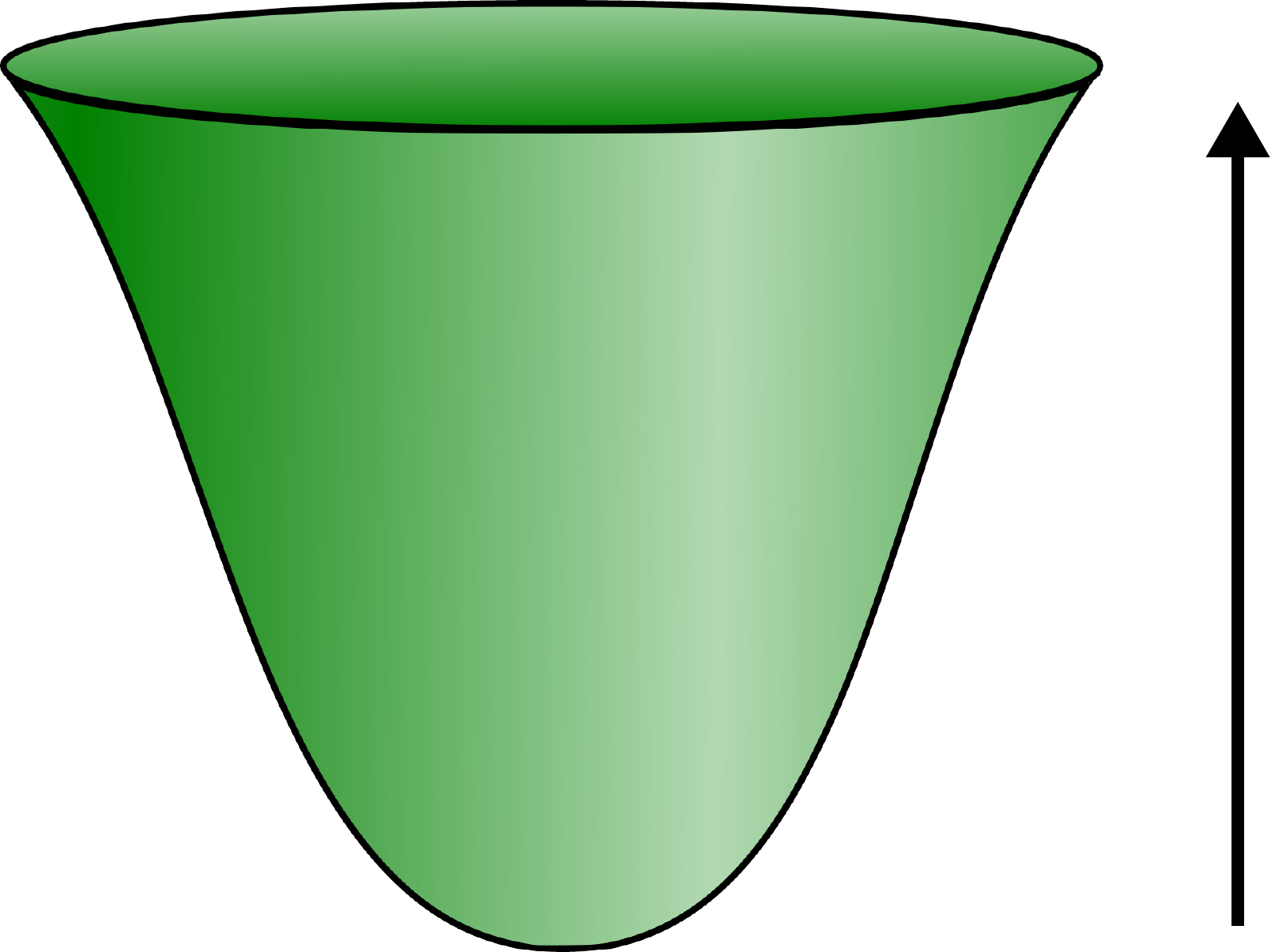}
\put (101,35) {$r$}
\put (94,-5) {IR}
\put (93,70) {UV}
\put (37,6) {{${\mathbf{R}^{p+2}}$}}
\end{overpic}
\caption{\label{greencap}The regular geometries interpolate between flat Euclidean D$p$-branes in the UV and $\mathbf{R}^{p+2}$ in the IR.}
\end{figure}
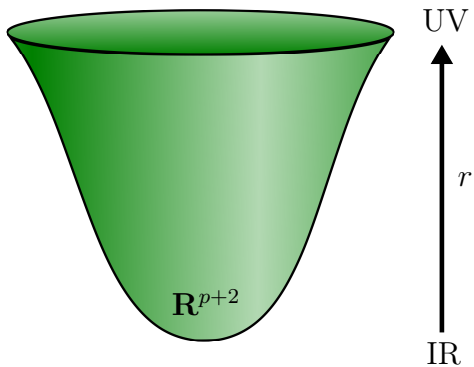

In the UV, i.e. for large values of $r$, the solution should asymptotically approach the flat brane domain wall solution \eqref{flatdomainwall} as depicted in Figure \ref{greencap}. Notice that in this UV limit  one finds $X=1$ and $Y=0$. In the IR region the scalar fields should approach a constant finite value in order to have a regular solution. These IR values for the scalars can be found as the critical points of the superpotential ${\cal W}$ (or $\widetilde{\cal W}$)
\be
\partial_\lambda {\cal W} =  D_\tau {\cal W}=0~,
\ee
which in terms of the new variables $X,Y$ read:
\be
\begin{array}{lll}\label{eq:XYIR} X_\text{IR}= \f{p}{3}~,\quad& Y_\text{IR}=\pm\f{2(p-3)}{3}~,&\text{for $p<3~$}\,,\\
X_\text{IR} = \f{p}{(6-p)(p-2)}~,\quad& Y_\text{IR} = \pm\f{2(p-3)}{(6-p)(p-2)}~,&\text{for $p>3~$}\,.\end{array}
\ee
The upper sign in the expressions above refers to a critical point of ${\cal W}$ whereas the lower sign refers to a critical point of $\widetilde{\cal W}$. Notice that for $p=4$ the critical value of the superpotential is at the UV point $X=1$. We will discuss this in more detail below. Even though $X$ and $Y$ approach fixed values in the IR, the scalar $\lambda$ can take any value $\lambda = \lambda_\text{IR}$. As discussed in Section~\ref{sec:holography} below, $\lambda_\text{IR}$ is related to the effective gauge coupling constant of the dual SYM theory at the IR energy scale set by the radius of the sphere.

Finally we want to point out that when solving the BPS equations in \eqref{lambdaeq}-\eqref{Aeq2} it sometimes proves useful to use the scalar $X$ as a new radial variable. This is possible since $X$ is a monotonic function of the radial variable $r$ in \eqref{eq:genpmetAnsatz}.

\subsection{Uplift to ten dimensions}
\label{subsec:uplift}

After this uniform treatment of the gauged supergravity theories in $p+2$ dimensions and their spherical brane solutions, we provide general uplift formulae that we use to obtain the spherical brane solutions in ten dimensions. These are distilled from the literature and brought into a universal form in Appendix \ref{App:sugra}. In this section we merely quote the results. The ten-dimensional metric takes the expected form in \eqref{metricansatz10d}
\be\label{eq:10dmetricgeneral}
\dd s_{10}^2 = \f{\e^{\lambda}}{\sqrt{Q}}\left(\dd s_{p+2}^2 + \f{\e^{\f{2(p-3)}{6-p}\lambda}}{g^2}\left(\dd \theta^2 + P\cos^2\theta~\dd \widetilde{\Omega}_2^2 + Q\sin^2\theta~\dd \Omega_{5-p}^2\right)\right)\,.
\ee
The squashing functions $P$ and $Q$ are determined in terms of the gauged supergravity scalars as
\bea\label{eq:PQdef}
P &=& \left\{\begin{array}{ll} X\big(X\sin^2\theta+(X^2-Y^2)\cos^2\theta\big)^{-1} & \text{for $p<3$}\,,\\
X\big(\cos^2\theta+X\sin^2\theta\big)^{-1} & \text{for $p>3$}\,,\end{array}\right.\\
Q &=& \left\{\begin{array}{ll} X\big(\sin^2\theta+X\cos^2\theta\big)^{-1} & \text{for $p<3$}\,,\label{eq:Qdef}\\
X\big(X\cos^2\theta+(X^2-Y^2)\sin^2\theta\big)^{-1} & \text{for $p>3$}\,.\end{array}\right.
\eea
The ten-dimensional dilaton is
\be\label{10Ddilaton}
\e^{2\Phi} = g_s^2\e^{\f{p(7-p)}{6-p}\lambda}~P~ Q^{\f{1-p}{2}}\,,
\ee
and the non-vanishing type II form fields are given by
\begin{equation}\label{eq:NSNSRRgenp}
\begin{aligned}
B_2 &=\e^{\f{p}{6-p}\lambda}\f{YP}{g^2 X}\cos^3\theta~\vol_2\,,\\
C_{5-p} &= i\e^{-\f{p}{6-p}\lambda}\f{YQ}{g_s g^{5-p}X}\sin^{6-p}\theta~\vol_{5-p}\,,\\
C_{7-p} &= \f{i}{g_sg^{7-p}}\left(\omega(\theta)+P\cos\theta~\sin^{6-p}\theta\right)\vol_2\w\vol_{5-p}\,.
\end{aligned}
\end{equation}
Here $\vol_{5-p}$ and $\vol_2$ refer to the volume forms on $\dd \Omega_{5-p}^2$ and $\dd \widetilde{\Omega}_2^2$, respectively, see \eqref{metricansatz10d} and \eqref{eq:metdS2}. The function $\omega(\theta)$ is defined such that in the UV the derivative of $C_{7-p}$ simply gives the volume form on the $(8-p)$--dimensional de Sitter space, namely
\be
\f{\dd}{\dd\theta}\left(\omega(\theta) + \cos\theta~\sin^{6-p}\theta\right) = (7-p)\cos^2\theta~\sin^{5-p}\theta\,.
\ee
%

\section{Details of the solutions}
\label{sec:Physics}

In this section we do a case-by-case study of the spherical brane solutions. The simplest example is provided by the near-horizon geometry of Euclidean D3-branes. It is simply given by ${\bf H}^5\times\text{dS}_5$. Writing the metric on ${\bf H}^5$ in global coordinates 
\be
\dd s^2_{{\bf H}^5} = \dd \eta^2 + \sinh^2\eta~ \dd \Omega_4^2~,
\ee
makes it clear that flat Euclidean D3-branes are described by the same supergravity solution as spherical D3-branes. This is of course a reflection of the fact that the worldvolume ${\cal N}=4$ SYM theory is conformal and the four-sphere is a conformally flat manifold. As discussed in Section \ref{sec:fieldtheory} placing non-conformal maximal SYM theories on spheres in a supersymmetric way must be accompanied by adding particular mass terms in addition to the standard conformal coupling term to the Lagrangian. In the bulk supergravity solutions this is manifested by modifying the usual flat Euclidean D$p$-brane solutions to new solutions of supergravity which we exhibit explicitly below.

\subsection{D6-branes}
\label{subsec:D6}

As discussed in Section \ref{subsec:Sugrap+2}, the case of spherical D6-branes is a degenerate limit of our equations since now we only have two scalar fields instead of three. This is consistent with the fact that for the maximal seven-dimensional SYM theory on $S^7$ the R-symmetry is unbroken. In addition we showed in \eqref{eq:Wp=6} that the superpotential is a purely imaginary constant which implies that the pseudoscalar $\chi$ does not appear in the scalar potential. This in turn means that the BPS equations derived in Appendix \ref{appB:D6} only result in first order equations for the scalar $\beta$ and the warp factor $A$. A first order equation for $\chi$ is obtained directly from the equations of motion. We refer to Appendix \ref{appB:D6} for further details on the eight-dimensional supergravity and the derivation of the BPS equations. Keeping this in mind it is still useful to mimic the structure of the BPS equations with $p<6$ and introduce new scalar variables
\be
\tau = i(X+Y)~,\quad \tilde\tau = -i(X-Y)~.
\ee
In these variables the BPS equations together with the equation of motion for $Y$ reduce to the following system of coupled first order ODEs
\bea\label{eq:p=6floweqns}
(X')^2 &=& \f94 g^2 X + 36{\cal R}^{-2} \e^{-14A}X^4~,\\
Y' &=& 6{\cal R}^{-1}X^2\e^{-7A}~,\label{p6Yeq}\\
(A')^2 &=& \f{1}{16}g^2 X^{-1} +{\cal R}^{-2}\e^{-2A}~,
\eea
where by prime we denote a derivative with respect to $r$. We have checked that this system of equations implies the equations of motion of the gauged supergravity theory. To solve the flow equations in \eqref{eq:p=6floweqns} it is convenient to use the metric function $A$ as a radial variable. One then finds
\be
X = \e^{6A}~.
\ee
We then proceed by defining yet another radial coordinate
\be
\rho =\text{arcsinh}\left( \f{4}{g{\cal R}}\e^{2A}\right)\,,
\ee
such that the metric takes the form
\be\label{p6sugrametric}
\dd s_8^2 = \f{g {\cal R}^3}{16} \sinh\rho\left(\dd \rho^2 + 4\dd\Omega_7^2\right)~,
\ee
and
\begin{equation}
X =\big((g{\cal R}/4) \sinh\rho\big)^3~.
\end{equation} 
The full solution of the gauged supergravity theory is obtained by integrating the equation
\be
\f{\dd Y}{\dd\rho}= \f{3}{16} g^3{\cal R}^3\sinh^3\rho~\tanh^2\rho~.
\ee
We will not need the lengthy analytic expression for $Y(\rho)$ in our analysis and thus refrain from presenting it here.

Many of the interesting properties of this solution are only apparent when uplifted to ten or eleven dimensions. Unfortunately we are not able to directly use the general formulae presented in Subsection \ref{subsec:uplift} since they are not valid for $p=6$. Nevertheless, since our eight-dimensional solution is rather simple the uplift formulae of \cite{AlonsoAlberca:2003jq} can be readily applied and yield the following type IIA background
\bea\label{eq:D6IIAsol}
\dd s_{10}^2 &=& \f{{\cal R}^2\e^{2\Phi/3}}{g_s^{2/3}}\left(\f14\dd \rho^2 + \dd \Omega_7^2 + \f{1}{16}\sinh^2\rho ~\dd \widetilde\Omega_2^2\right)~,\\
H_3 &=& \f{3}{g^2g_s^2}\e^{2\Phi}\dd\rho\w \vol_2~,\\
F_2 &=& \f{i}{g_sg}\vol_2~,\\
\e^{2\Phi} &=& g_s^2\left(\f{g{\cal R}}{4}\sinh\rho\right)^3~.
\eea
For $\rho \to 0$ the metric and dilaton approach that of a D6-brane in flat space:
\bea
\dd s_{10}^2 &\approx& \f{1}{\sqrt{H}}{\cal R}^2\dd\Omega_7^2 + \sqrt{H}\left(\dd \tilde r^2 + \tilde r^2 \dd \widetilde\Omega_2^2\right) ~,\\
\e^{2\Phi} &\approx& \left(H\right)^{-3/2}~,
\eea
where $16\tilde r = g({\cal R}\rho)^2$ and $H = 1/g\tilde r$. The function $H$ is precisely the harmonic function for $N$ D6-branes in the near-horizon limit upon replacing $g$ with $N$ using \eqref{gsugra}
\be
H = \f{g_s N \ell_s}{2\tilde r}~.
\ee
We thus conclude that $\rho \to 0$ should be identified with the UV limit of the dual gauge theory. In the limit $\rho\to \infty$ we should be exploring the IR regime of the field theory where the finite size of $S^7$ should play a role. Indeed the eight-dimensional metric \eqref{p6sugrametric} in this limit caps off in the expected regular manner wheras the dilaton blows up. This is an indication that we must further uplift our type IIA solution to eleven dimensions and interpret it in M-theory. The uplift to eleven dimensions has to be done with some care because the two-form field strength $F_2$ and its one-form potential, $C_1$, are imaginary. The one-form potential appears in the eleven-dimensional metric as a Kaluza-Klein vector and so if it is imaginary it would render the eleven-dimensional metric complex. This is resolved by remembering that, as discussed in Section~\ref{sec:flatDp}, our type II solutions can be interpreted as solutions of Hull's type II$^*$  theories \cite{Hull:1998vg}. Hull has argued in \cite{Hull:1998ym} that the type IIA$^*$ theory uplifts to an eleven-dimensional version of M-theory with two time directions called M$^*$-theory. In our formulation this means that we can apply the standard uplift formulae presented in Appendix \ref{App:conventions} with a purely imaginary M-theory circle parametrized by $i\omega$ with $\omega\in{\bf R}$. Doing this for the solution in \eqref{eq:D6IIAsol} we obtain, quite surprisingly, a metric on ${\bf H}^{2,2}/{\bf Z}_N\times S^7$ where ${\bf H}^{2,2} \equiv \SO(3,2)/\SO(2,2)$. Explicitly we find
\be
\dd s_{11}^2 = \f{{\cal R}^2}{4{g_s^{2/3}}}\left(\dd s_{4}^2  + 4~\dd \Omega_7^2\right)~,
\ee
where
\be
\dd s_{4}^2=\dd\rho^2 -  \f{1}{4}\sinh^2\rho \left(\dd t^2 - \cosh^2t~\dd\psi^2 + (N^{-1}\dd \omega - \sinh t ~\dd \psi)^2\right)~,
\ee
is a metric on ${\bf H}^{2,2}$ with three-dimensional anti-de Sitter spacetime boundary, albeit in the wrong signature. Even though the coordinate $\omega$ is timelike, it should still be treated as periodic, just like in the standard relation between type IIA and eleven-dimensional supergravity. We have parametrized the M$^*$-theory circle such that $\omega$ has periodicity $\omega\sim\omega+4\pi$. Notice that crucially the metric on AdS$_3$ is not regular unless $N=1$. In fact the structure of the metric is precisely that of an extremal BTZ black hole. The analytic continuation of this metric to eleven spacelike dimensions yields a metric on
\be
{\bf H}^4/{\bf Z}_N\times S^7~,
\ee
where the four-dimensional hyperbolic space has a boundary that is a Lens space $S^3/{\bf Z}_N$. Given that the eleven-dimensional metric above is closely related to the standard AdS$_4\times S^7$ solution of eleven-dimensional supergravity, it is not surprising to find that up to factors of $N$, the four-form flux is the standard one
\be\label{eq:D6M2G4flux}
G_4=\f{3i}{2g^2g_s^2\ell_s}\left(\f{g{\cal R}}{4}\sinh\rho\right)^3\dd\rho\w \vol_2\w\dd\omega = \f{3i}{L_4}\vol_{{\bf H}^{2,2}}~,
\ee
where we have introduced 
\be\label{eq:L4D6def}
L_4= \f{\cal R}{2 g_s^{1/3}}~.
\ee
As expected we also find that the M2-brane flux
\be\label{eq:NM2D6def}
N_\text{M2} = \f{1}{(2\pi \ell_s)^6i}\int G_7 =\f{2 L_4^6}{\pi^2 \ell_s^6}\in {\bf Z}\,,
\ee
is properly quantized. The explicit appearance of $i$ in \eqref{eq:D6M2G4flux} and \eqref{eq:NM2D6def} is a result of our choice of conventions for IIA$^*$ and M$^*$ theories. We will discuss the holographic interpretation of this curious eleven-dimensional background further in Section~\ref{sec:holography}.

\subsection{D5/NS5-branes}
\label{subsec:D5}

For $p=5$ the solution is constructed in the maximal $\SO(4)$ gauged supergravity in seven dimensions \cite{Samtleben:2005bp} which arises as a consistent truncation of type IIB supergravity on $S^3$, see \cite{Boonstra:1998mp}. Just like the well known $\SO(5)$ gauged theory \cite{Pernici:1984xx}, obtained by reducing eleven-dimensional supergravity on $S^4$, this theory has maximal supersymmetry. 

\begin{figure}[h!]
\centering
\begin{overpic}[width=0.75\textwidth,tics=2]{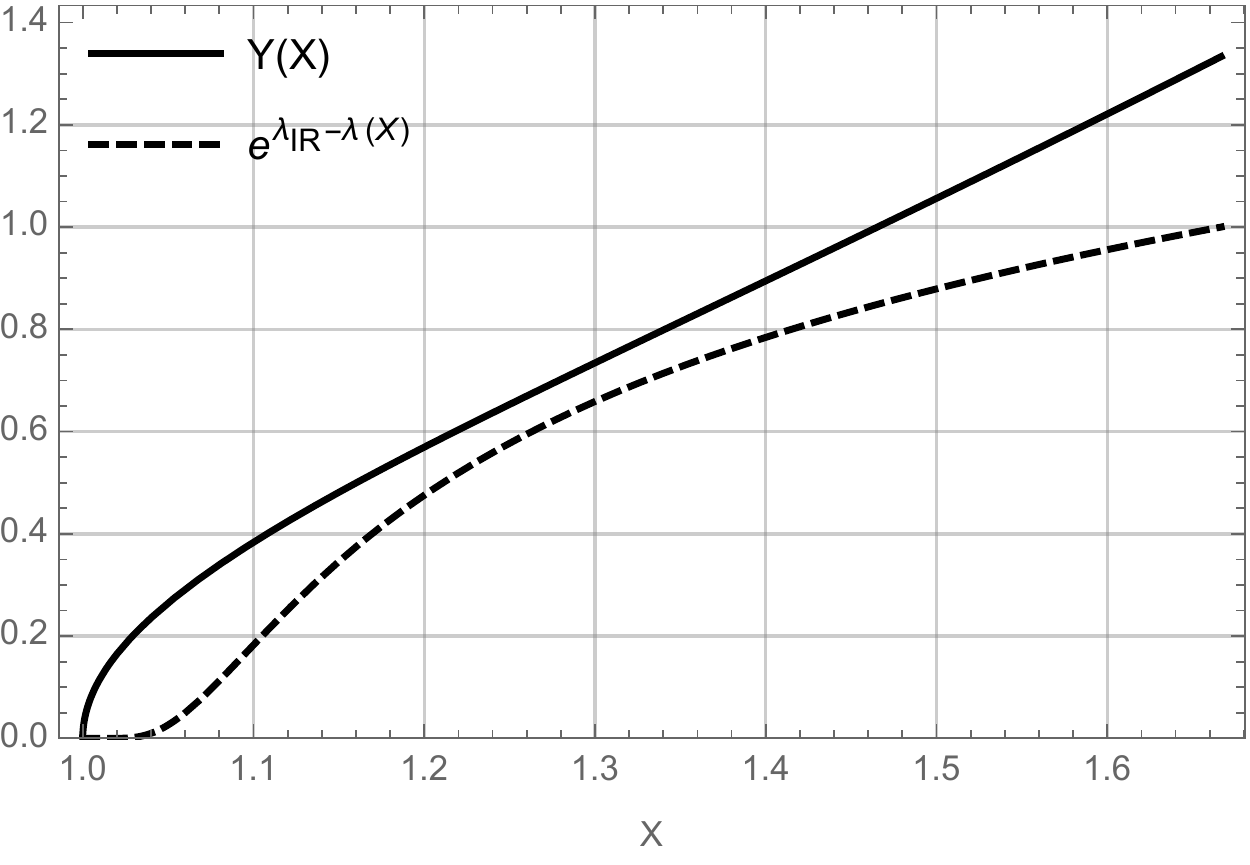}
\put (97,63.6) {\Large$\bullet$}
\put (97,49.8) {\Large$\bullet$}
\end{overpic}
\caption{\label{fig:D5flow}A numerical solution for the functions $Y(X)$ and $\lambda(X)$ in the case $p=5$. Notice that since \eqref{D5eq} is quadratic in $Y$, the function $-Y$ is also a solution with $\lambda$ unchanged. The far UV region is at $Y=0$, $X=1$ and the IR region where the $S^6$ smoothly caps of is indicated by the solid dots.}
\end{figure}

Using the three scalar truncation of the $\SO(4)$ theory discussed in Appendix \ref{appB:D5} one can derive the BPS equations given in \eqref{lambdaeq}-\eqref{Aeq2}. As mentioned in the previous section it is convenient to use $X$ as a coordinate and express the remaining scalar fields $Y$ and $\lambda$ as functions of $X$. The solutions to the BPS equations are then fully determined by the following ODE
\be\label{D5eq}
\f{\dd Y^2}{\dd X} = \f{Y^2}{X}\left(\f{1-16X+15X^2-9Y^2}{2-8X+6X^2-3Y^2}\right)~,
\ee
together with the integral
\be\label{D5lambdaeq}
\lambda(X) =\lambda_\text{IR} + \int_{X_\text{IR}}^X \f{3~\dd x}{5\big(1-x\big)}\left[\Big(x-\f{1}{3}\Big)\f{\dd \log Y(x)}{\dd x}-1\right]~.
\ee
In terms of the new coordinate, the seven-dimensional metric takes the form
\be
\dd s_{7}^2 = \f{(1-3X)^2-9Y^2}{g^2X\e^{4\lambda}}\left(\f{\dd X^2}{\left(2-8X+6X^2-3Y^2\right)^2}+\f{X^2}{Y^2}\dd \Omega_{6}^2\right)~.
\ee
Unfortunately we have not been able to find an analytic solution to the equations above that connects the IR values of $X$ and $Y$ given in \eqref{eq:XYIR} to the UV values $X=1$, $Y=0$. To construct such a solutions to \eqref{D5eq}, we therefore have to resort to numerical methods. As will become quite familiar when solving the supergravity BPS equations for various values of $p$, the solution for $Y$ as a function of $X$ is completely fixed by the IR boundary conditions \eqref{eq:XYIR}. The only physical integration constant that appears in the solution is the value of the scalar $\lambda$ in the IR and conveniently $\lambda_\text{IR}$ only appears as an additive constant in the integral expression \eqref{D5lambdaeq}. A numerical plot of the solution is given in Fig. \ref{fig:D5flow}. Its uplift to a solution of type IIB supergravity is given by the formulae in Section~\ref{subsec:uplift} with $p=5$.

As discussed above, the numerical solution interpolates between a regular IR region where $X$, $Y$, and $\lambda$ approach a constant value and the metric caps off smoothly and the near-horizon geometry of D5-branes in the UV. The uplift to ten dimensions provides a full solution to the type II supergravity equations of motion describing D5-branes wrapped on $S^6$. Furthermore by an $\SL(2,{\bf R})$ transformation we can obtain a solution describing NS5-branes (or more generally $(p,q)$-fivebranes) wrapped on $S^6$ and all equations of motion of course remain satisfied. The wrapped NS5-brane solution is particularly interesting since the sphere provides an IR cut-off of the linear dilaton geometry sourced by NS5-branes in flat space. As briefly discussed in Section \ref{sec:fieldtheory}, the UV completion of SYM theories in six dimensions is believed to be given by a non-local, non-gravitational theory called little string theory (LST). This theory can be understood as the decoupling limit of NS5-branes  where the string coupling vanishes $g_s\to 0$ \cite{Seiberg:1997zk}. A holographic model for LST was considered in \cite{Aharony:1998ub} and studied in more detail in \cite{Giveon:1999px,Giveon:1999tq}. The original construction is based on the linear dilaton vacuum which is simply the near-horizon limit of $N$ flat NS5-branes.\footnote{This background can be obtained from \eqref{nearhorizonbranesmetric} for $p=5$ by an $\SL(2,{\bf R})$ transformation.} The metric and dilaton of this type IIB solution are
\be
\dd s_{10}^2 = \dd s_6^2+\dd\eta^2 + g^{-2}\dd \Omega_3^2~,\qquad \Phi = \log g_s - g \eta~,
\ee
where $g^2 = g_s N^{-1} \ell_s^{-2}$. Although these fields together with a Yang-Mills instanton provide an exact background of heterotic string theory \cite{Callan:1991at}, interpreting it in the context of holography is somewhat problematic due to the singular bahaviour of the dilaton for large negative $\eta$. In particular, it makes the holographic computation of LST correlation functions impossible without further information about the singular region $\eta\to -\infty$ \cite{Aharony:1998ub}. In \cite{Giveon:1999px,Giveon:1999tq} a resolution of the singularity was proposed whereby the $N$ NS5-branes are spread out on a circle breaking the $\SO(4)$ isometry group of the space transverse to the branes to an $\SU(2)$ subgroup. In a T-dual frame the singularity corresponds to the origin of an ALE space
\be\label{ALEeq}
z_1^N + z_2^2 + z_3^2 =0~,
\ee
and the resolution of the singularity is achieved by introducing a non-zero right-hand side in \eqref{ALEeq}. Our type IIB supergravity solution provides an alternative way to resolve the problem. Remember that the singularity can be understood as a result of the dual SYM theory becoming weakly coupled in the IR. As we have explained, placing the SYM on $S^6$ introduces an effective IR cut-off set by the radius of the sphere. In the supergravity description this is manifested by the smooth cap-off of the geometry in the IR. More explicitly we find that the dilaton of our spherical NS5-brane background takes the form
\be
\Phi = \log \left(\f{g_s\sqrt{P}}{X}\right) - 5\lambda~,
\ee
which in the IR reduces to 
\be
\Phi_\text{IR} = \log \left(\f{3g_s}{\sqrt{5(3\sin^2\theta + 5\cos^2\theta)}}\right) - 5\lambda_\text{IR}~,
\ee
and therefore $\e^\Phi$ can be made arbitrarily small throughout the full solution by suitably tuning $\lambda_\text{IR}$. It will be most interesting to understand better this spherical NS5 background and its implications for the physics of LST.

\subsection{D4-branes}
\label{subsec:D4}

Our solutions for spherical D4-branes are constructed in a six-dimensional gauged supergravity theory obtained by reducing the maximal $\SO(5)$ gauged supergravity in seven dimensions on a circle. As explained in more detail in Appendix~\ref{appB:D4}, we first introduce a seven-dimensional scalar $x$ that breaks $\SO(5)\to \SU(2)\times \U(1)$ together with a $\U(1)$ gauge field  ${\cal A}$. Reducing this theory on a circle introduces the dilaton $\phi$  as well as an additional scalar field arising from the component of the gauge field on the reduction circle, i.e. ${\cal A}= \chi \dd\omega$ where $\omega$ is the coordinate on the circle. As a result we obtain the desired three scalar fields, $x$, $\phi$ and $\chi$.\footnote{These scalars are the ones discussed in Section~\ref{subsec:Sugrap+2}.} After rewriting the BPS equations with the scalar $X$ as a coordinate the system reduces to a single ODE which controls the full solution 
\be\label{D4eq}
\f{\dd Y^2}{\dd X} = \f{Y^2(1-12X+12X^2-4Y^2)}{2X(1-X)(1-2X)}~.
\ee
This equation is solved by 
\be\label{p4analytic}
Y^4 = cX(1-X)\Big((1-2X)^2-4Y^2\Big)^2~,
\ee
where $c$ is an integration constant. The critical point of the superpotential determines the IR values of the scalar field as in \eqref{eq:XYIR} which for $p=4$ yields $X_\text{IR}=1$ and $Y_\text{IR}=\pm 1/2$. However, the analytic solutions \eqref{p4analytic} only reach the IR for diverging $c$, i.e. when $(1-2X)^2-4Y^2=0$. This  is a solution to the BPS equation \eqref{D4eq} but it is not physical since the metric
\be
\dd s_{6}^2 = \f{(1-2X)^2-4Y^2}{g^2X\e^{\lambda}}\left(\f{\dd X^2}{4(1-2X)^2(X-1)^2}+\f{X^2}{Y^2}\dd \Omega_{5}^2\right)~,
\ee
vanishes completely. All solutions in \eqref{p4analytic} with finite $c$ correspond to gravitational domain walls with singular IR behavior. These singular flows still provide solutions to the ten-dimensional equations of motion via the uplift formulae in Section~\ref{subsec:uplift}. Furthermore an uplift of these solutions to eleven-dimensional supergravity is given in Section~\ref{appB:D4}. Still, the conclusion remains that there is no smooth solution with running scalar $X$ that connects the UV to a regular IR. This is perhaps not surprising since the IR value of the scalar $X$ is located at $X=1$ which is also the UV value for $X$.

We are thus lead to explore solutions with constant $X=1$. The original BPS equations \eqref{lambdaeq}-\eqref{Aeq2} are solved by $2Y = \e^{2\lambda_{\rm IR}-2 \lambda}$ where $\lambda_{\rm IR}$ is an integration constant. The BPS equation for $\lambda$ then reduces to 
\be
(\lambda')^2 = \f{g^2}{16}\e^{-5\lambda}\left(\e^{4\lambda}-\e^{4\lambda_{\rm IR}}\right)~.
\ee
Notice that the scalar $\chi=\e^{2\lambda_{\rm IR}}/2$ is constant (cf. \eqref{taudef} and \eqref{tautoXY}). 
Using $\lambda$ as a coordinate, the six-dimensional metric can be written as
\be
\dd s_6^2 = \f{8 \e^{3\lambda -2\lambda_{\rm IR}}}{g^2}\left(\sinh^{-1}(2\lambda-2\lambda_{\rm IR})\dd \lambda^2 + \sinh(2\lambda-2\lambda_{\rm IR})\dd\Omega_5^2\right)~.
\ee
Since the six-dimensional supergravity theory used to construct this solution is obtained from a reduction of the maximal seven-dimensional $\SO(5)$ gauged supergravity it is possible to uplift the solution above to seven dimensions. Performing this uplift, see Appendix~\ref{appB:D4}, one finds that the metric and the scalar fields in seven dimensions are simply that of the maximally supersymmetric AdS$_7$ (or rather $\mathbf{H}_7$) vacuum of the gauged supergravity, albeit with an asymptotic $S^5\times S^1$ metric on the boundary. There is however also a non-vanishing gauge field $\mathcal{A} = \chi \dd\omega$, see \eqref{eq:metAD4app}, which is pure gauge since the six-dimensional scalar field $\chi$ is constant. Note that due to the topology of $S^1$ it requires a large gauge transformation to set the field $\mathcal{A}$ to zero.

The six-dimensional spherical D4-brane solution above can also be uplifted to ten-dimensional type IIA supergravity. The explicit form of the solution is 
\bea
\dd s_{10}^2 &=& \f{\e^\lambda}{\sqrt{Q}}\left(\dd s_6^2 + \f{\e^\lambda}{g^2}\left(\dd \theta + \cos^2\theta\, \dd \tilde\Omega_2^2 + Q \sin^2\theta\, \dd \zeta^2 \right)\right)~,\\
\e^{2\Phi} &=& g_s^2 \e^{6\lambda} Q^{-3/2}~,\\
B_2 &=& \f{\e^{2\lambda_{\rm IR}}}{2g^2}\cos^3\theta\, \vol_2~,\\
C_1 &=& \f{i \e^{2\lambda_{\rm IR}}}{2g_s g}\e^{-4\lambda}Q\sin^2\theta\,\dd\zeta~,\\
C_3 &=&-\f{i}{g_s g^3}\cos^3\theta~\vol_2\w\dd\zeta~.
\eea
This background is of the general form discussed in Section~\ref{subsec:uplift} with
\begin{equation}
P=1~,\qquad\qquad Q= 4 \left(4 - \e^{4\lambda_{\rm IR}-4\lambda} \sin^2\theta\right)^{-1}~.
\end{equation}
In the IR region the scalar $\lambda$ which determines the behavior of the dilaton is finite and approaches the constant $\lambda_{\rm IR}$. In the UV region however, the scalar $\lambda$ diverges and the the type IIA dilaton blows up. This indicates that the proper description of the solution is in eleven-dimensional supergravity. To find this eleven-dimensional background we can use the uplift formulae in Appendix \ref{App:conventions}. However, just like in Section \ref{subsec:D6}, we should remember that we are working in the type IIA$^{*}$ theory of Hull which uplifts to the M$^*$-theory in which the eleven-dimensional circle is timelike. We take this into account by using a pure imaginary $x_{11}$. Keeping this in mind we find the following eleven-dimensional background
\bea\label{eq:11dM*uplift}
\dd s_{11}^2 &=& \f{1}{g_s^{2/3}g^2} \Bigg(8\e^{2\tilde\lambda}\left(\f{\dd\tilde\lambda^2}{\sinh 2\tilde\lambda} + \sinh 2\tilde\lambda~\dd\Omega_5^2\right) - 4 \e^{4\tilde \lambda} \dd \omega^2\nonumber\\
&& \qquad\qquad+\, \dd \theta^2 + \cos^2\theta \dd \tilde\Omega_2^2 + \sin^2\theta (\dd \omega-\dd\zeta)^2\Bigg)~,\\
A_3 &=&\f{i}{g_s g^3}\cos^3\theta~(\dd \omega-\dd\zeta)\w \vol_2~,
\eea
where we shifted $\tilde\lambda = \lambda-\lambda_\text{IR}$ and the eleventh direction is parametrized by $g g_s \e^{2\lambda_{\rm IR}} x_{11}/2=i\omega$. This eleven-dimensional solution is valid in the limit when $\tilde\lambda$ is very large. As it turns out the first line of \eqref{eq:11dM*uplift} is simply the global metric on AdS$_7$ whereas the second line is a metric on four-dimensional de Sitter space. Indeed the full solution is the analytic continuation of the well-known AdS$_7\times S^4$ solution of standard eleven-dimensional supergravity to the M$^*$-theory.
%
%
It is encouraging to find that in the far UV region of our spherical D4-brane solution we find the metric in \eqref{eq:11dM*uplift} which should be associated with the near-horizon limit of Euclidean M5-branes. This is in line with the expectation discussed in Section~\ref{sec:fieldtheory} that the five-dimensional maximally symmetry SYM theory on $S^5$ flows to the superconformal $(2,0)$ theory on $S^5\times S^1$ in the UV.

\subsection{D2-branes}
\label{subsec:D2}

\begin{figure}[h!]
\centering
\begin{overpic}[width=0.75\textwidth,tics=2]{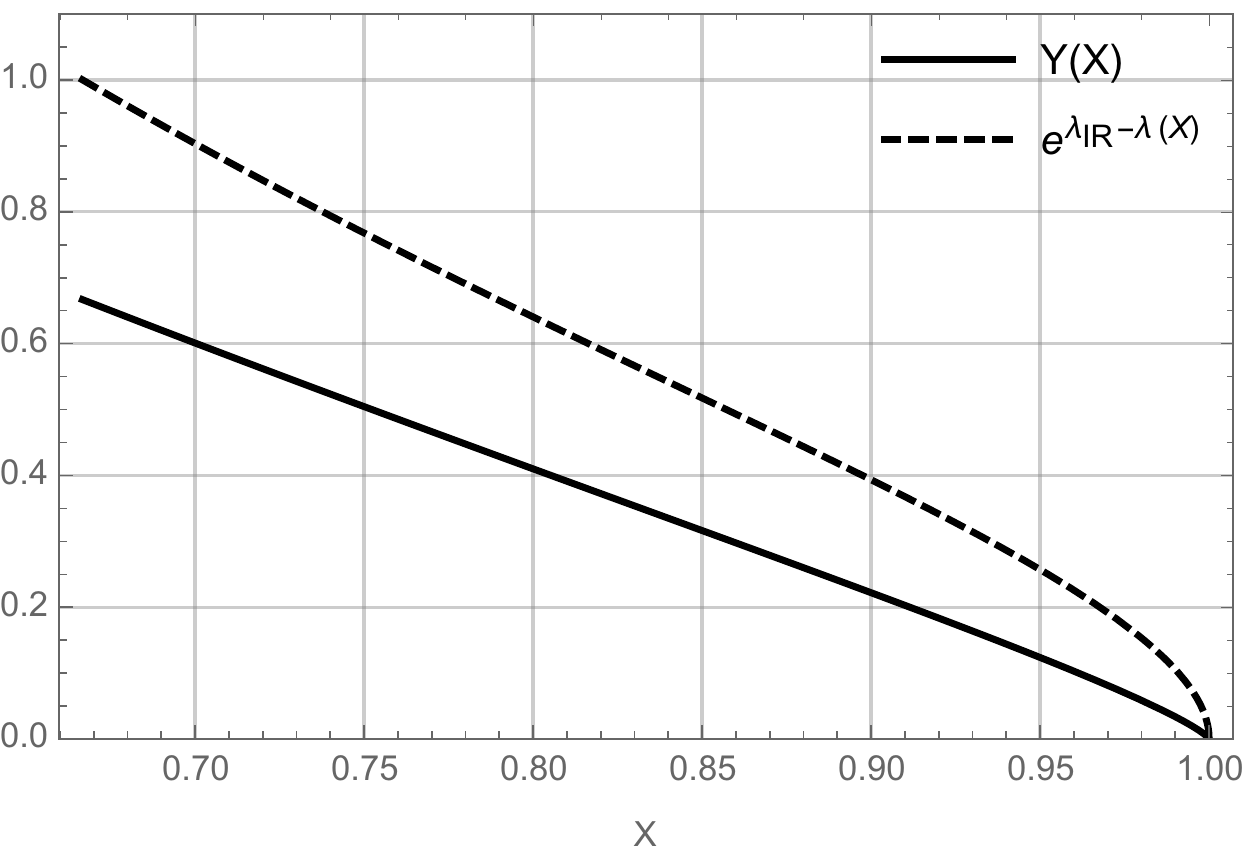}
\put (5.5,61.3) {\Large$\bullet$}
\put (5.5,43.8) {\Large$\bullet$}
\end{overpic}
\caption{\label{fig:D2flow}The full analytic solution for the functions $Y(X)$ and $\lambda(X)$ in the case $p=2$. The UV region is at $X=1$, $Y=0$ and the IR region is indicated by the solid dots.}
\end{figure}

Spherical D2-branes are constructed in maximal supergravity in four dimensions with $\ISO(7)$ gauge symmetry. This theory was first constructed by Hull in \cite{Hull:1984yy} and later argued to arise as a  consistent truncation of type IIA supergravity on $S^6$ \cite{Hull:1988jw}, see Appendix~\ref{App:D2} for more details. Using once again the scalar $X$ as radial coordinate we can reduce the set of BPS equations \eqref{lambdaeq}-\eqref{Aeq2} to a single ODE
\be
\f{\dd Y^2}{\dd X} = \f{Y^2(7X^2-4X-Y^2)}{X(2X(X-1)+Y^2)}~.
\ee
This ODE is solved by 
\be\label{p2analytic}
Y^4 = cX\Big(X(X-1)-Y^2\Big)^3~,
\ee
where $c$ is an integration constant. After setting $c=-1$ we obtain a solution connecting the UV values of the scalars $X=1$, $Y=0$ with their IR values as in \eqref{eq:XYIR} with $p=2$. This choice of integration constant still leaves us with six distinct solutions for $Y(X)$. However, two of them, $Y=\pm X$, are not physical since the metric
\be\label{eq:p2sugrametric}
\dd s_{4}^2 = \f{X^2-Y^2}{g^2X \e^{-\lambda/2}}\left(\f{\dd X^2}{(2X(X-1)+Y^2)^2}+\f{X^2}{Y^2}\dd \Omega_{3}^2\right)~
\ee
vanishes for these flows. Of the remaining four solutions, only two flow to the (regular) IR. These are given by
\be\label{p2Yfinalsol}
Y^2 = \f{(1-X)}{2X}\left((1-X)(1+2X) +\sqrt{(1-X)(1+3X)}\right)~.
\ee
The BPS equation for $\lambda$ can now be readily integrated and yields
\be
\e^{2(\lambda-\lambda_\text{IR})} = 1-X+\f{Y^2}{X}~,
\ee
with $Y^2$ given by \eqref{p2Yfinalsol}. To illustrate this analytic solutions we plot it in Figure~\ref{fig:D2flow}.

This four-dimensional supergravity solution can be uplifted to a ten-dimensional solution of type IIA supergravity using the uplift formulae in Section \ref{subsec:uplift} with $p=2$. We have verified that all equations of motion in ten dimensions are satisfied by the above solution. In the UV one finds that $X \to 1$ and $Y\to 0$ and the ten-dimensional solution reduces to the near-horizon limit of D2-branes with flat worldvolume. Although the four-dimensional solution is completely regular, see \eqref{eq:p2sugrametric}, the ten-dimensional background appears to be singular in the IR due to the fact that the metric function $Q$ as defined in \eqref{eq:Qdef} blows up for $X=2/3$ and $\theta=0$. This problem can be circumvented completely by a double analytic continuation
\be\label{eq:analyticcontinuationpless3}
\theta \to \pi/2 + i\tilde\theta~,\qquad \psi\to i\tilde\psi~,
\ee
where $\theta$ appears in the uplift formula for the metric \eqref{eq:10dmetricgeneral} and $\psi$ is a coordinate on the dS$_2$ in \eqref{eq:metdS2}. The analytic continuation \eqref{eq:analyticcontinuationpless3} leaves the functions $P$ and $Q$ positive in the full range of the new coordinates $0\le\tilde\theta\le\infty$ and $2/3\le X \le 1$ and so the metric as well as the other fields are now regular. Furthermore the metric remains with signature (1,9) where the $\tilde \theta$ parametrizes the time direction. A careful study of the form-fields shows that the continued solution is also a solution of type IIA$^*$ with all R-R fields pure imaginary and NS-NS fields real. Finally, the global symmetry of the solution matches with the field theory expectation \eqref{eq:isomgroup} where $\SU(1,1)$ is now realized as the isometry group of the hyperbolic plane spanned by $(t,\tilde\psi)$. The need to perform the analytic continuation in \eqref{eq:analyticcontinuationpless3} can perhaps be traced to the fact that in the SYM Lagrangian in \eqref{symlagrangian} the coefficient of one of the bosonic bilinear terms changes sign as one goes from $p>3$ to $p<3$. This interpretation is also consistent with the fact that we have to perform the same analytic continuation for the supergravity solution describing spherical D1-branes.

As discussed in Section~\ref{sec:flatDp}, flat D2-branes are singular in the IR since the dilaton blows up. It is well-known that this singularity can be interpreted by going to eleven-dimensional supergravity since flat D2-branes uplift to M2-branes smeared over the M-theory circle. The IR singularity can therefore be understood as a direct consequence of the smearing and its resolution is achieved by replacing the smeared M2-branes by a point-like stack localized on the circle. In this way the singular supergravity solution is resolved by replacing it with the AdS$_4\times S^7$ solution of eleven-dimensional supergravity. On the gauge theory side this interpretation is mirrored by the expectation that maximal SYM theory in three dimensions flows to the conformal ABJM theory in the deep IR. For our spherical D2-brane solution there is no singularity in the IR and in fact the dilaton is never so large as to warrant an uplift to eleven dimensions. In the dual field theory the interpretation is clear - placing three-dimensional maximal SYM on a three-sphere introduces an IR cut-off and the RG flow never reaches the superconformal ABJM theory in the IR. As a final comment we note that the spherical D2-brane supergravity solution should lie in region (b) of Figure 1 in \cite{Itzhaki:1998dd}.

\subsection{D1-branes}
\label{subsec:D1}

The final example we consider is the supergravity solution for spherical D1-branes. In this case we have to deviate slightly from our general approach of first finding the solution of interest in a lower-dimensional gauged supergravity and then uplifting it to ten dimensions. The reason for this is that we are not aware of an appropriate three-dimensional supergravity theory that is obtained by a consistent truncation of type IIB supergravity on $S^8$. Nevertheless we are still able to make progress and find the solution directly in ten dimensions by solving a system of ODEs, which resemble BPS equations derived from a three-dimensional supergravity theory, and are obtained by analytically continuing the equations in Section~\ref{subsec:Sugrap+2} to $p=1$. We then use the solution of these effective BPS equations in a ten-dimensional background of the form presented in Section~\ref{subsec:uplift} with $p=1$ and check explicitly that the equations of motion of type IIB supergravity are obeyed. It is highly non-trivial that this procedure works and we consider this sufficient evidence the resulting solution describes the backreaction of spherical D1-branes.

To describe the solution we use again the scalar $X$ to parametrize the radial direction. The BPS equation for $Y$ then reduces to
\be\label{D1eq}
\f{\dd Y^2}{\dd X} = \f{Y^2(7X^2+1-Y^2)}{X(2(X^2-1)+Y^2)}~.
\ee
A particular solution to this ODE that interpolates between the IR at $(X,Y)=(1/3,\pm4/3)$ and the UV at $(X,Y)= (1,0)$ is
\be
Y^2 = \f{(X+1)(1-X)^2}{X}~.
\ee
Next, the equation for $\lambda$ can be readily integrated,
%
%
resulting in 
\be\label{D1lambdaeq}
\lambda = \lambda_{\rm IR} + \f{5}{2}\log\f{1-X}{2X}~.
\ee 
The three-dimensional metric in \eqref{eq:10dmetricgeneral} is explicitly given in terms of the scalars $Y$ and $\lambda$ by
\be
\dd s_{3}^2 = \f{(1+X)^2-Y^2}{g^2X\e^{-\f{4}{5}\lambda}}\left(\f{\dd X^2}{(2(1+X)(X-1)+Y^2)^2}+\f{X^2}{Y^2}\dd \Omega_{2}^2\right)~.
\ee
One can then set $p=1$ in the formulae in Section~\ref{subsec:uplift} and obtain an explicit solution of type IIB supergravity.

\begin{figure}[h!]
\centering
\begin{overpic}[width=0.75\textwidth,tics=2]{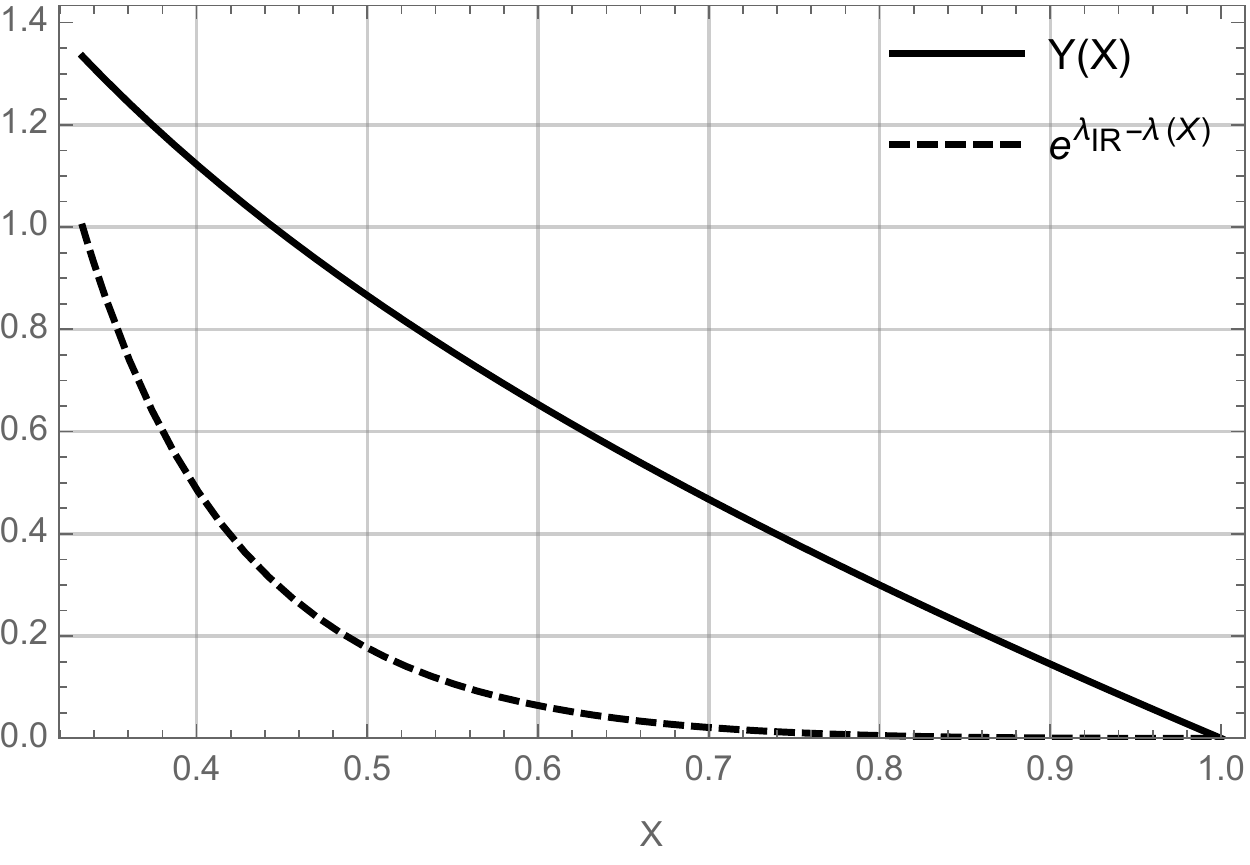}
\put (5.3,63.2) {\Large$\bullet$}
\put (5.3,50.0) {\Large$\bullet$}
\end{overpic}
\caption{\label{fig:D1flow}A analytic solution in the case $p=1$. The UV region is at $X=1$, $Y=0$ and the IR is indicated by the solid dots.}
\end{figure}

Again we note that regularity in the IR completely fixes the profile for $Y$ as a function of $X$ and the only integration constant of the solution is the one that appears in the expression for $\lambda$ in \eqref{D1lambdaeq}. A plot of the analytical solution for the scalar fields is given in Figure \ref{fig:D1flow}. As was the case for the other spherical brane solutions above we find that in the UV region the ten-dimensional background we construct is asymptotic to the flat D1-brane solution of type IIB supergravity. In the IR region the three-dimensional metric caps off smoothly which reflects the IR cut-off provided by the scale of the $S^2$ in the dual two-dimensional maximal SYM theory. The ten-dimensional metric on the other hand exhibits a similar feature as the D2-brane solution in that there is a region in the plane spanned by $X$ and $\theta$ for which the metric function $Q$ becomes negative. Just as for the spherical D2-brane solution we can cure this problem by performing the analytic continuation \eqref{eq:analyticcontinuationpless3}. This renders the new ten-dimensional configuration a completely regular background of type IIB$^*$ theory.

\section{Holographic interpretation}
\label{sec:holography}

It is natural to conjecture that the supergravity backgrounds presented in Sections \ref{sec:SpherDp} and \ref{sec:Physics} are holographically dual to the maximal SYM theory on $S^{p+1}$ described in Section~\ref{sec:fieldtheory}. So far we have presented some basic evidence for this duality --- the global symmetries and the supersymmetry of the gauge theory and the supergravity solutions agree. More refined tests of this duality are harder to perform since for $p\neq3$ the SYM theory is not conformal and thus in general we do not have an asymptotically AdS$_{p+2}$ region in the geometry. This in turn implies that we cannot rely on the standard holographic dictionary which is well-developed for asymptotically AdS space-times. Nevertheless some progress can be made and our goal in this section is to evaluate the on-shell action of the supergravity solutions and compare the result with the free energy of the SYM theory on $S^{p+1}$ computed by supersymmetric localization in \cite{Minahan:2015jta,Minahan:2015any}. 

It is important to emphasize that we are interested in the scaling of the free energy of the SYM theory with the rank of the gauge group, $N$, and the effective dimensionless 't~Hooft coupling $\lambda_\text{eff}$ which is defined by
\be\label{lambdaeffdef}
\lambda_\text{eff}(E) = g_\text{YM}^2 N E^{p-3}~,
\ee
see for example \cite{Itzhaki:1998dd}. Here $E$ is the energy scale in the field theory at which the coupling is defined. In the supersymmetric localization computation in \cite{Minahan:2015jta,Minahan:2015any}, with which we want to compare our supergravity results, the energy scale is set by the inverse radius of the sphere. It should also be noted that in this field theory calculation there is an implicit choice of regularization scheme. The relation between the energy scale in a non-conformal SYM theory and the ``radial coordinate'' in a  dual supergravity background is a subtle problem that has been discussed in a similar context in \cite{Peet:1998wn,Boonstra:1998mp,Kanitscheider:2008kd}. We will make use of these results in the calculations below. 

In standard AdS holography the holographic computation of the free energy $F$ is performed by evaluating the renormalized on-shell action of the gravitational background. For our non-AdS backgrounds the procedure of holographic renormalization is less developed and the calculation is more subtle. It is perhaps possible to use the results of \cite{Kanitscheider:2008kd} to approach this problem in a more systematic fashion however, since here we are mainly interested in the scaling of $F$ with $N$ and $\lambda_\text{eff}$, we will circumvent going through the holographic renormalization procedure. 

As discussed in Section \ref{subsec:uplift}, the $p+2$-dimensional supergravity action in \eqref{sugraaction} is obtained by a dimensional reduction of type II ten-dimensional supergravity on a $(8-p)$ dimensional de Sitter space equipped with a squashed metric. For the purposes of evaluating the on-shell action we analytically continue all our solutions to be fully Euclidean such that the de Sitter part of the metric becomes a sphere. This is needed in order to obtain a finite Newton constant for the $(p+2)$-dimensional effective supergravity theory whose on-shell action we are interested in computing. The Newton constant in $p+2$ dimensions, $\kappa_{p+2}^2$, is obtained by directly integrating out the $(8-p)$-dimensional sphere  starting from the ten-dimensional action. Ignoring numerical volume factors coming from the integration over the sphere we find that $\kappa_{p+2}^2$ is related to the fundamental string theory parameters by
\be
2\kappa_{p+2}^2 = \f{2\kappa_{10}^2 g_s^2}{\text{vol}_{8-p}}  \sim (2\pi \ell_s)^8 g_s^2 g^{8-p}~.
\ee

It turns out that to determine the scaling of the free energy with $N$ and $\lambda_\text{eff}$ it is sufficient to find how the on-shell action scales with the integration constant $\lambda_{\rm IR}$. To this end we should find the scaling of the integrand in the action in \eqref{sugraaction} with the metric function $\e^A$, which in turn depends on $\lambda_\text{IR}$ through its dependence on $\lambda$. This dependence is determined by rewriting the algebraic relation \eqref{eq:algebraicA} in the variables $(X,Y)$ using \eqref{tautoXY}. We then make use of the fact that in these variables the profiles for $X$ and $Y$ do not involve the integration constant $\lambda_\text{IR}$ and in fact $\lambda_\text{IR}$ only appears as an additive constant to an explicit integral expression, cf. \eqref{D5lambdaeq} and \eqref{D1lambdaeq}. This means that we can easily determine the scaling of the metric function with $\lambda_\text{IR}$:
\be
{\cal R}\e^{A} \sim g^{-1}\e^{\f{p-3}{6-p}\lambda_\text{IR}}~.
\ee
It should be noted that for the supergravity solutions corresponding to flat D$p$-branes the integration constant $\lambda_\text{IR}$ is not physical since it can be removed by shifting the radial coordinate, see for example \eqref{flatdomainwall}. For the spherical D$p$-brane backgrounds, however, $\lambda_\text{IR}$ is physical since the shift of the radial coordinate can be used to either fix the radial location of the IR region or to fix the value of $\lambda$ in the IR but not both simultaneously.

Combining all the pieces together one can show that all terms in the integrand of the action in \eqref{sugraaction} scale in the same way with $\e^\lambda$ which results in the following scaling of the on-shell action, or alternatively the supergravity dual of the free energy in the dual field theory,
\be\label{eq:Fsugrascaling}
F_{\text{sugra}} \sim \int\dfrac{({\cal R}\e^{A})^p}{2\kappa_{p+2}^2}  \sim \f{\e^{\f{p(p-3)}{6-p}\lambda_\text{IR}}}{(2\pi \ell_s g)^8 g_s^2}~.
\ee
This expression is schematic since the on-shell action formally diverges due to the integration over the infinite range of the radial coordinate and the asymptotic behavior of the metric function $\e^{A}$ in the UV region. These divergences should be regularized by a proper application of a holographic renormalization procedure for non-AdS space-times. Nevertheless, we believe that the scaling of the on-shell action presented on the right hand side of \eqref{eq:Fsugrascaling} is the correct one. This is due to the fact that the standard local holographic counterterms one can write down appear with the same power of $\e^{\lambda_{\rm IR}}$ as in \eqref{eq:Fsugrascaling}.

At this point we need to relate the supergravity parameters $g$, $\lambda_\text{IR}$, and $g_s$ to quantities in the dual field theory. First we use the relation between $g$ and the rank of the gauge group $N$ in \eqref{gsugra}. The relation between the effective 't~Hooft coupling and the ten-dimensional dilaton was written down in \cite{Kanitscheider:2008kd} (see also \cite{Peet:1998wn,Boonstra:1998mp})
\be\label{lambdaeff}
\e^\Phi = \f{1}{N} c_d \lambda_\text{eff}^{\f{7-p}{2(5-p)}}~,
\ee
where $c_d$ is a numerical constant which can be found in eq. $(2.21)$ of \cite{Kanitscheider:2008kd} but is not important for the scaling argument we are making. The relation \eqref{lambdaeff} implicitly relates the energy scale of the field theory appearing in \eqref{lambdaeffdef} to the radial coordinate of the supergravity solution through the radial dependence of the dilaton. Combining this with \eqref{10Ddilaton}, we find a relation between the supergravity parameter $\lambda_\text{IR}$ and the effective 't~Hooft coupling\footnote{Some care has to be taken when combining \eqref{lambdaeff} with \eqref{10Ddilaton} since the expression in \eqref{10Ddilaton} depends on both the internal and radial coordinates. To arrive at \eqref{eq:lambdaIRlambdaeff} we used \eqref{10Ddilaton} evaluated in the UV region. } 
\be\label{eq:lambdaIRlambdaeff}
\e^{\f{p(7-p)}{2(6-p)}\lambda_\text{IR}}\sim \f{1}{g_s N}\lambda_\text{eff}^{\f{7-p}{2(5-p)}}~.
\ee
We are now in a position to express the holographic free energy in \eqref{eq:Fsugrascaling} in terms of field theory quantities and find
\be\label{Fscalinglaw}
F_{\text{sugra}} \sim N^2 \lambda_\text{eff}^{\f{p-3}{5-p}}~.
\ee
This supergravity result nicely agrees with the field theory calculation done in \cite{Minahan:2015any} where the free energy of large $N$ maximal SYM theory on $S^{p+1}$ was obtained by supersymmetric localization for $2<p<5$  and analytically continued to general $p\ne 5$. Notice that for $p=5$ the relation in \eqref{lambdaeff} degenerates and thus the result in \eqref{Fscalinglaw} is not reliable. This was also observed in the context of flat D$p$-brane supergravity solutions in \cite{Peet:1998wn,Boonstra:1998mp} and is perhaps due to the fact that the world-volume theory for D5/NS5-branes is not an ordinary local QFT in the UV. Note also that although we have focused on the non-conformal SYM theories for $p\neq3$ the result in \eqref{Fscalinglaw} nicely reproduces the well-known scaling with $N$ of the free energy of the $\mathcal{N}=4$ SYM theory.

The case of spherical D4-branes is somewhat special since in the far UV region the maximal five-dimensional SYM theory flows to the superconformal $(2,0)$ theory in six dimensions which is holographically dual to AdS$_7\times S^4$. It is well-known that the free energy of this SCFT scales as $N^3$. We can recover this scaling by combining \eqref{Fscalinglaw} with $p=4$ and \eqref{lambdaeffdef}. Notice that as discussed in \cite{Kallen:2012zn} the holographic evaluation of the AdS$_7$ on-shell action disagrees, by a constant factor of order 1, with the large $N$ result for the SYM free energy on $S^5$ computed by supersymmetric localization. It will be very interesting to resolve this discrepancy and understand the role played by the asymptotically  AdS$_7$ solution discussed in Section~\ref{subsec:D4}.

The calculation above is not valid for $p=6$ since, as discussed in Section~\ref{subsec:Sugrap+2}, the scalar $\lambda$ is not part of the $p+2$-dimensional supergravity theory. Fortunately there is an alternative method that can be used to calculate the holographic free energy of the spherical D6-brane solution. As discussed in Section~\ref{subsec:D6} the eleven-dimensional (Euclidean) supergravity description of the spherical D6-brane is given by $S^7\times {\bf H}^4/{\bf Z}_N$. We can calculate the free energy of this background by compactifying on $S^7$, treat the resulting background as a  solution of supergravity in four dimensions and evaluate its on-shell action. This brings us into the familiar territory of holographic renormalization for asymptotically locally AdS$_4$ (or ${\bf H}^4$ in Euclidean signature) spaces. The on-shell action for the ${\bf H}^4/{\bf Z}_N$ background of interest can be evaluated after adding the usual Gibbons-Hawking and curvature counter terms. To this end we apply the results of  \cite{Emparan:1999pm} to find the holographic free energy\footnote{Here we use $16\pi G_N=2\kappa_{4}^2$ to denote Newtons constant of the four-dimensional supergravity theory.}
\be\label{freeenergyforD6}
F = \f{L_{4}^2}{4\pi G_N}\f{2\pi^2}{N}~,
\ee
where the factor of $N$ appears because the ${\bf Z}_N$ action reduces the volume of the boundary manifold $S^3/{\bf Z}_N$ as compared to the volume of $S^3$. As usual, $G_N$ is the four-dimensional Newton constant:
\be
G_N = \f{3\pi^3\ell_s^9}{8L_{4}^7}~.
\ee
The free energy can then be expressed in terms of $N_\text{M2}$ as well as the number of D6-branes $N$ and takes the form
\be
F = \f{\pi L_4^2}{2 N G_N} = \f{\sqrt{2}\pi}{3N} N_\text{M2}^{3/2}~.
\ee
%
$N_\text{M2}$ has a somewhat mysterious interpretation in the seven-dimensional field theory. Notice that $N_\text{M2}$ is a dimensionless quantity built from the radius of the seven-sphere. It is thus natural to conjecture that it is related to the dimensionless 't~Hooft coupling of the field theory. Indeed if we take ${\cal R}^{-1}$ as the energy scale in the effective 't~Hooft coupling \eqref{lambdaeffdef} we find 
\be\label{eq:lambdaeffNM2}
2^3\pi^6\lambda_\text{eff}^{-2}({\cal R}^{-1}) = 2^3\pi^6(g_\text{YM}^2N {\cal R}^{-3})^{-2} = \f{2^3\pi^6{\cal R}^6}{(2\pi)^8 g_s^2\ell_s^6N^2} = \f{N_\text{M2}}{N^2}~.
\ee
To arrive at this expression we have used \eqref{eq:L4D6def} and \eqref{eq:NM2D6def}. This result is consistent with the relation in \eqref{lambdaeff}. Using this we find the final expression for the holographic free energy of spherical D6-branes 
\be\label{eq:FD6final}
F = \f{2^5\pi^{10}N^2}{3\lambda_\text{eff}^{3}} ~.
\ee
It is quite nice to see that this expression follows from analytically continuing the one in \eqref{Fscalinglaw} to $p=6$ and therefore agrees with the field theory results in \cite{Minahan:2015any}. It should be stressed that the numerical coefficients in \eqref{eq:FD6final} have been computed using a ``holographic renormalization scheme'', i.e. using the relation in \eqref{eq:lambdaeffNM2}. It will be most interesting to compute this numerical coefficient from the field theory and compare with the holographic result in \eqref{eq:FD6final}. Finally, we would like to point out that it was observed in \cite{Minahan:2015any} that the localization calculation for the path integral of seven-dimensional maximal SYM on $S^7$ exhibits features reminiscent of a theory with conformal symmetry. Clearly this QFT is not conformal so this feature appears puzzling. It is tempting to speculate that this localization observation is related to the fact that in our supergravity solution the local isometries of ${\bf H}^4/{\bf Z}_N$ close the Euclidean three-dimensional conformal algebra $\SO(4,1)$.

\subsection{Thermal free energy}
\label{Sec:thermalfree}

As we have emphasized numerous times, the finite size of $S^{p+1}$ provides an IR cut-off for the low-energy dynamics of the SYM theory which is compatible with supersymmetry. A more commonly used IR cut-off is to consider  SYM  at finite temperature. This of course breaks supersymmetry but nevertheless offers the possibility for a qualitative comparison with the holographic results above. The supergravity dual description of a $(p+1)$-dimensional maximal SYM theory at finite temperature is given by a $(p+2)$-dimensional black brane solution which we summarize below.

The black branes of interest are most easily described as solutions to the $(p+2)$-dimensional gauged supergravity theory described in Section~\ref{subsec:Sugrap+2}. In contrast to the spherical brane solutions however these nonsupersymmetric backgrounds preserve the full gauge symmetry. This fits well with the fact that in the dual gauge theory at finite temperature the R-symmetry is preserved. The metric of the solution takes a standard black-hole form
\be
\dd s_{p+2}^2 = \dd r^2 + \e^{2A(r)}\left(-h(r)\dd t^2 + \dd x_p^2\right)~,
\ee
where $\dd x_p^2$ is the metric on $\mathbf{R}^p$. In addition to the metric the only field with a non-trivial profile is the scalar $\lambda(r)$. The equations of motion reduce to the following set of equations\footnote{These black brane solutions are clearly well-known and studied in many references, see for example \cite{Peet:2000hn}. For convenience we rederive them here in our conventions and notation.}
\bea
A &=& {\f{9-p}{(6-p)(p-3)}\lambda}~,\\
\log(1-h) &=& \f{2p(p-7)}{(6-p)(p-3)}(\lambda-\lambda_0)~,\\
\left(\e^{\f{p-3}{6-p}\lambda}\right)' &=& -\f{g(3-p)^2}{2p}\sqrt{h}~,
\eea
where $\lambda_0$ is an integration constant. When the integration constant is chosen such that $h=1$ we recover the supersymmetric flat domain wall solution in \eqref{flatdomainwall}. The horizon of the black hole is located where $\lambda\to\lambda_0$ and the asymptotic infinity (UV) is located where $h\to1$. Notice that for $p<3$ the UV is located for large negative $\lambda$ but for $p>3$ it is located at large positive $\lambda$. In the near-horizon region the metric takes the form
\be
\dd s_{p+2}^2 = \dd r^2 -\f{g^2(7-p)^2}{4}\e^{\f{2(5-p)p}{(6-p)(p-3)}\lambda_0}(r-r_0)^2\dd t^2 + \e^{\f{2(9-p)}{(6-p)(p-3)}\lambda_0}\dd x_p^2~.
\ee
The temperature of the black hole can be determined using the standard trick of ensuring that the near-horizon metric does not have conical singularities when analytically continued to Euclidean time, $it$. The result is
\be
T = \f{g(7-p)}{4\pi}\e^{\f{(5-p)p}{(6-p)(p-3)}\lambda_0}~.
\ee
This black brane solution can be uplifted to ten dimensions using the formulae in Section \ref{subsec:uplift} and the metric in string frame reads:
\be
\dd s_{10}^2 =  (gU)^{\f{p-7}{2}}\left( h^{-1}\dd U^2 + (gU)^{7-p}\left(-h~\dd t^2 + \dd x_p^2\right) + U^2\dd\Omega_{8-p}^2\right)~,
\ee
where $gU = \e^{\f{2p}{(6-p)(p-3)}\lambda}$ such that
\be	
h = 1-\f{U_0^{7-p}}{U^{7-p}}~,\qquad gU_0 = \e^{\f{2p}{(6-p)(p-3)}\lambda_0}~.
\ee
The dilaton and R-R fields are the same as for the flat supersymmetric brane solution in \eqref{nearhorizonbranesdilaton} and \eqref{nearhorizonbranesC}. The effective 't~Hooft coupling can be easily computed using \eqref{lambdaeff} and \eqref{nearhorizonbranesdilaton} and is given in terms of the temperature 
\be\label{eq:lambdaeffT}
\lambda_\text{eff} \sim \left(g_s N\right)^{\f{2(5-p)}{7-p}} \left(\f{T}{g}\right)^{p-3}\sim g_s N \left(2\pi \ell_s T\right)^{p-3}~.
\ee
An alternative way to arrive at the same relation for $\lambda_\text{eff}$ is to first identify the energy scale in the QFT, $E$, with the temperature of the black hole, $T$. Using this and the relation in \eqref{lambdaeffdef} one obtains again \eqref{eq:lambdaeffT}. The entropy of the black branes was computed in \cite{Itzhaki:1998dd} and can be evaluated in terms of the Einstein frame area of the horizon which is 
\be
{\cal A}_\text{Einst} = \e^{-2\Phi(U_0)} {\cal A}_\text{str} \sim g_s^{-2} \left( gU_0\right)^{\f{9-p}{2}}g^{p-8}V_p~,
\ee
where $V_p$ is the spatial volume of the D$p$-branes. The area of the horizon determines the entropy of the black brane via $S = 2{\cal A}_\text{Einst}/\kappa_{10}^2$ which in terms of the field theory quantities takes the form
\be
S \sim N^2 \lambda^{\f{p-3}{5-p}}_\text{eff} ~T^{p}V_p~.
\ee
It is then clear that the thermal free energy $F\sim TS$ has the same scaling in terms of $N$ and $\lambda_\text{eff}$ as the supersymmetric free energy in \eqref{Fscalinglaw}. This can be viewed as another consistency check of our field theory interpretation of the spherical brane backgrounds as holographically dual to the maximal SYM theory on $S^{p+1}$.

\section{Conclusions}
\label{sec:conclusions}

In this paper we constructed explicit supergravity solutions which preserve sixteen real supercharges and describe the backreaction of spherical D$p$-branes with $1\leq p\leq 6$. We also argued that these backgrounds are holographically dual to the planar limit of the maximally supersymmetric Yang-Mills theory on $S^{p+1}$. An immediate consistency check of this claim is provided by the fact that the global symmetries and supersymmetry of the gauge theory and the supergravity solutions are the same. As a more refined check of the duality we showed that the on-shell action of the supergravity solutions agrees with recent results from supersymmetric localization about the planar limit of the free energy of the SYM theory. There are several interesting questions for further research that arise from these spherical brane solutions which we discuss briefly below.

While in Section~\ref{sec:holography} we described a way to extract holographic information from the on-shell action of the spherical brane solutions and compared that successfully with supersymmetric localization results, it is clear that the procedure we employed is not rigorous. It is certainly desirable to understand better how to apply holographic renormalization for the spherical brane solutions in order to be able to systematically extract supergravity results for correlation functions in the dual SYM theory on $S^{p+1}$. The results of \cite{Kanitscheider:2008kd} can perhaps be adapted and used in this context. Once the holographic renormalization procedure is under control one can study other gauge theory observables using the spherical brane solutions. A natural candidate are supersymmetric Wilson line operators. Depending on the representation of the gauge group these operators should be described by probe fundamental strings or D-branes and their expectation value is captured by the on-shell action of the probe. It would also be interesting to study these line operators in the dual gauge theory using supersymmetric localization and perhaps establish additional checks of this non-conformal holographic duality. The holographic calculation of the expectation value of  a BPS Wilson loop in the fundamental representation for the non-conformal four-dimensional $\mathcal{N}=2^*$ theory on $S^4$ was recently performed in \cite{Bobev:2018hbq} and the result agrees with supersymmetric localization. This gives us reasons to be optimistic that a similar calculation for the spherical D$p$-brane solutions above should be within reach.

We constructed our ten-dimensional spherical brane solutions by first studying supersymmetric domain walls of lower-dimensional maximal gauged supergravity and making use of uplift formulae. There should be a more systematic way to construct brane solutions with curved world volumes directly in ten- and eleven-dimensional supergravity. There is of course the standard Maldacena-N\'u\~nez construction of supergravity solutions sourced by branes with curved worldvolumes which preserve supersymmetry via a partial topological twist \cite{Maldacena:2000mw,Maldacena:2000yy}, see also \cite{Gauntlett:2003di} for a review. However it is well-established that supersymmetric gauge theories on curved manifolds can preserve supersymmetry in more general ways \cite{Festuccia:2011ws}. It is thus natural to ask whether and how the branes of string and M-theory realize these alternative supersymmetric gauge theories. This question was recently discussed in \cite{Triendl:2015lka,Minasian:2017rgh} but it is fair to say that the subject deserves a deeper and more systematic exploration.

Given that maximal SYM is believed to possess a unique Lagrangian on $S^{p+1}$ it is natural to conjecture that our spherical brane solutions are the unique ones preserving 16 supercharges. This in turn should provide the unique IR cut-off of the dual SYM theory with this amount of supersymmetry. However it should be possible to find a generalization of our construction to gauge theories with smaller amounts of supersymmetry by adding suitable couplings to the SYM Lagrangian. Recently this was explored in field theory using supersymmetric localization in \cite{Minahan:2015any,Gorantis:2017vzz}. The supergravity description of such a construction should bear similarities with the $\mathcal{N}=1^*$ and $\mathcal{N}=2^*$ mass deformations of the four-dimensional $\mathcal{N}=4$ SYM on $S^4$ which were recently explored in a holographic context in \cite{Bobev:2013cja,Bobev:2016nua,Bobev:2018hbq}. It will also be interesting to construct similar supersymmetric supergravity solutions dual to the maximal SYM theory on different curved manifolds. Perhaps a natural example to consider is given by a squashed $S^{p+1}$ when $p=2k$ is an even integer. In this case we can view the sphere as a $\U(1)$ bundle over $\mathbf{CP}^k$ and squash the Einstein metric while preserving $\SU(k)\times \U(1)$ invariance. This construction should preserves 8 supercharges and the partition function of the field theory should be computable by supersymmetric localization, see for example \cite{Hama:2011ea,Imamura:2012xg,Imamura:2012bm}. It will be most interesting to explore these questions further.

Finally we should stress that the description of spherical branes in this work is through the supergravity solutions describing their backreaction on space-time. It will certainly be interesting to have a better understanding of the Euclidean D$p$-branes with spherical worldvolume from the perspective of open string theory. The proper framework for such a study appears to be the type II$^{*}$ string theories introduce by Hull \cite{Hull:1998vg,Hull:1998fh,Hull:1998ym}. It is important to understand better the role of these somewhat exotic variations of string theory and how to microscopically describe Euclidean D$p$-branes with a curved worldvolume.

\section*{Acknowledgements}

We thank Iosif Bena, Adam Bzowski, Adolfo Guarino, Joe Minahan, Anton Nedelin, Leonardo Rastelli, Kostas Skenderis, and Thomas Van Riet for useful discussions. The work of NB and PB is supported in part by an Odysseus grant G0F9516N from the FWO. FFG is a Postdoctoral Fellow of the Research Foundation - Flanders. We are also supported by the KU Leuven C1 grant ZKD1118 C16/16/005.

\appendix


\section{Conventions for type II supergravity}
\label{App:conventions}

The spherical D$p$-brane backgrounds constructed in this paper solve the equations of motion of ten-dimensional type II supergravity. This theory comes in two flavors, type IIA and type IIB, depending on whether the chirality of the supersymmetry generators $\epsilon_{1,2}$ is opposite or the same. The bosonic field content consists of the NS-NS sector, common to both type IIA and type IIB, and the R-R sector. The metric $G_{MN}$, the dilaton $\Phi$, and the three-form $H_3$ build up the NS-NS sector, while the R-R sector contains the $n$-form field strengths $F_n$, with $n=0,2,4$ for type IIA and $n=1,3,5$ for type IIB. In type IIA $F_0$, i.e. the Romans mass, does not have any propagating degrees of freedom and is set to zero throughout this paper. In type IIB $F_5$ has to obey a self-duality condition. The fermionic field content consists of a doublet of gravitinos, $\psi_M$, and a doublet of dilatinos, $\lambda$. The components of these doublets are again of opposite chirality in type IIA and of the same chirality in type IIB.

In this paper we use the democratic formalism in which the number of R-R fields is doubled such that $n$ runs over $0,2,4,6,8,10$ for type IIA and $1,3,5,7,9$ for type IIB \cite{Bergshoeff:2001pv}. This redundancy is removed by introducing duality conditions for all R-R fields
\begin{equation}
F_n = (-1)^{\f{(n-1)(n-2)}{2}}\star_{10} F_{10-n} \,.
\end{equation}
These duality conditions should be imposed by hand after deriving the equations of motion from the action. The bosonic part of the action written in string frame is given by\footnote{We mostly follow the conventions of \cite{Blumenhagen:2013fgp}.}
\begin{equation}\label{actionII}
S_{{\rm bos}} = \f{1}{2\kappa_{10}^2}\int \star_{10}\Big[ \e^{-2\Phi}\Big( R + 4 |\dd \Phi|^2 - \f{1}{2} |H_3|^2 \Big) - \f{1}{4} \sum_{n}|F_n|^2 \Big]\,,
\end{equation}
where the ten-dimensional Newton constant $\kappa_{10}$ is related to the string length through $4\pi\kappa_{10}=(2\pi l_s)^8$ and we have defined
\begin{equation}
	\star_{10}|A|^2 \equiv \star_{10}\f{1}{n!} A_{\mu_1\dots\mu_n}A^{\mu_1\dots\mu_n} = \star_{10} A \wedge A \,.
\end{equation}
This action should be completed by its fermionic counterpart, which we do not write explicitly, and is invariant under the following supersymmetry variations\footnote{In these formulae we have implicitly chosen positive chirality spinors in type IIB supergravity.}
\begin{equation}\label{typeIIsusyvar}
	\begin{aligned}
	\delta\psi_M^1 &= \Big( \nabla_M - \f14 \slashed H_{3M} \Big)\epsilon^1 + \f{1}{16}\e^\Phi \sum_n \slashed {F_n} \Gamma_M\Gamma_{(10)}\epsilon^2 \, , \\
	\delta\psi_M^2 &= \Big( \nabla_M + \f14 \slashed H_{3M} \Big)\epsilon^2 - \f{1}{16}\e^\Phi \sum_n (-1)^{\frac{(n-1)(n-2)}{2}}\slashed {F_n} \Gamma_M\Gamma_{(10)}\epsilon^1 \, , \\
	\delta\lambda^1 &= \Big( \slashed\partial \Phi - \f12 \slashed H_3 \Big)\epsilon^1 + \f{1}{16}\e^\Phi \Gamma^M \sum_n \slashed {F_n} \Gamma_M\Gamma_{(10)}\epsilon^2 \, , \\
	\delta\lambda^2 &= \Big( \slashed\partial \Phi + \f12 \slashed H_3 \Big)\epsilon^2 - \f{1}{16}\e^\Phi \Gamma^M \sum_n (-1)^{\frac{(n-1)(n-2)}{2}}\slashed {F_n} \Gamma_M\Gamma_{(10)}\epsilon^1\,,
	\end{aligned}
\end{equation}
where $\Gamma_M$ are the ten-dimensional gamma matrices and $\Gamma_{(10)}$ is the chirality operator. The Feynman slash notation for an $n$-form field is defined as follows
\begin{equation}
	(\slashed{A}_n)_{M_{k+1}\cdots M_n} \equiv \Gamma^{M_1\cdots M_k}(A_n)_{M_1\cdots M_kM_{k+1}\cdots M_n}\,,
\end{equation}
for $k\leq n$ and $\Gamma^{M_1\cdots M_k} \equiv \f{1}{k!}\Gamma^{[M_1}\cdots \Gamma^{M_k]}$ is the totally antisymmetric product of $k$ gamma matrices.

The Bianchi identities and equations of motion derived from the action \eqref{actionII} are
\begin{equation}
	\dd H_3 =0\,, \qquad \text{and}\qquad 
	\dd (\e^{-2\Phi}\star_{10} H_3) + \frac{1}{2}\sum_n \star_{10} F_n\wedge F_{n-2} =0\,,
\end{equation}
for the NS-NS field $H_3$ and
\begin{equation}
	\dd F_n -H_3 \wedge F_{n-2} = 0\,,
\end{equation}
for the R-R form fields. The NS-NS and R-R fluxes can be written in terms of potentials as
\begin{equation}
F_n = \dd C_{n-1} - H_3 \w C_{n-3}\,,\qquad\qquad H_3 = \dd B_2\,.
\end{equation}
The dilaton and the Einstein equations of motion can be written as
\begin{equation}
\begin{aligned}
0 &= \nabla^2 \Phi - |\dd \Phi|^2 + \f14 R - \f18 |H_3|^2  \, ,\\
0 &= R_{MN} + 2 \nabla_M\nabla_N \Phi - \f12 |H_3|^2_{MN} - \f14 \e^{2\Phi} \sum_n |F_n|^2_{MN}  \,,
\end{aligned}
\end{equation} 
where we have defined
\begin{equation}
	|A_n|^2_{MN} \equiv \f{1}{(n-1)!}{(A_n)_{M}}^{M_2\cdots M_n}(A_n)_{NM_2\cdots M_n} \,.
\end{equation}
In the strong coupling limit, $g_s \gg 1$, type IIA string theory is expected to be described by M-theory. Therefore it will sometimes be useful to uplift our ten-dimensional type IIA supergravity solutions to eleven-dimensional supergravity. When compactified on a circle the eleven-dimensional theory has two parameters, the eleven-dimensional Newton constant $\kappa_{11}$ and the radius of the circle $R_{11}$. These are related to the ten-dimensional parameters as follows 
\begin{equation}
	R_{11}=\ell_s \qquad \text{and}\qquad \kappa_{11}^2 = 2\pi R_{11}\kappa_{10}^2 \,.
\end{equation}
The bosonic fields of eleven-dimensional supergravity are the metric and a three-form potential $A_3$. Their dynamics is governed by the following action
\begin{equation}
	S = \f{1}{2\kappa^2_{11}}\int \star_{11}\Big[ R-\f{1}{2}|\dd A_3|^2 \Big] - \frac{1}{12\kappa_{11}^2}\int A_3\w \dd A_3\w \dd A_3\,.
\end{equation}   
To reduce to ten dimensions we make the following Kaluza-Klein ansatz 
\begin{equation}\label{eq:KK11to10}
	\begin{aligned}
	\dd s_{11}^2 &= \e^{-\f23\Phi}G_{MN}\dd x^M dx^N + \e^{\f43\Phi}\left(\dd x^{11}+C_1\right)^2\,,\\
	A_3 &= C_3 + B_2\wedge \dd x^{11}\,.
	\end{aligned}
\end{equation}
All fields appearing on the right hand side of \eqref{eq:KK11to10} are the ten-dimensional type IIA fields in string frame.

\section{Flat Euclidean D-branes}
\label{App:flatEbranes}

In this appendix we explicitly show that flat Euclidean D$p$-branes are indeed supersymmetric solutions of type II$^*$ supergravity. For more details on these theories, including the type II$^*$ actions, see for example \cite{Bergshoeff:2007cg}.

The supersymmetry variations are exactly the same as those of regular type II supergravity, see \eqref{typeIIsusyvar}, with the only difference that the R-R fields now have to be treated as purely imaginary and the spinor obeys an unusual reality condition. For type IIA$^*$ the spinors satisfy a MW$^*$ condition
\be
\epsilon^* = -\mathcal{C} A \Gamma_{(10)}\epsilon\,,
\ee
and similarly for type IIB$^*$
\be
\epsilon^* = \mathcal{C} A\, \sigma_3\epsilon\;,
\ee
where $\epsilon = (\epsilon_1 , \epsilon_2)^T$ and $A$ and $\mathcal{C}$ define respectively Dirac conjugation, $\bar{\chi}^D \equiv \chi^\dagger A$, and Majorana conjugation, $\bar{\chi} \equiv \chi^T \mathcal{C}$. The reality conditions for the spinors are thus equivalent to
\be\label{relA}
\bar{\epsilon} = - \bar{\epsilon}^D \Gamma_{(10)}\,,
\ee
for type IIA$^*$ while for type IIB$^*$ we find
\be
\label{relB}
\bar{\epsilon} = \bar{\epsilon}^D \sigma_3\,.
\ee
We can now check explicitly that the flat Euclidean branes of Hull are indeed $\frac{1}{2}$-BPS solutions of type II$^*$ supergravity, i.e. they preserve 16 real supercharges. The solutions are given by
\bea
\dd s_{10}^2 &=& H^{-1/2}\dd s_{p+1}^2+H^{1/2}\dd s_{1,8-p}^2\, ,\\
\e^\phi &=& g_s H^{(3-p)/4}\, ,\\
C_{p+1} &=& i(g_s H)^{-1} \text{vol}_{p+1}
\eea
In these solutions $\dd s_{p+1}^2$ is the metric of flat $(p+1)$-dimensional Euclidean space, $\dd s_{1,8-p}^2$ is the Minkowski metric on $\mathbf{R}^{1,8-p}$, and $H$ is a harmonic function on this Minkowski space.

Inserting these solutions into the supersymmetry variations we see that they can indeed be solved by imposing the usual D$p$-brane projector with an extra $i$ inserted
\be\label{proj}
\left(1 + i \Gamma^{0\dots p}\Gamma_{(10)}\mathcal{P}\right)\epsilon = 0\,.
\ee
Here $\mathcal{P} = \sigma_1$ when $\f{p(p+1)}{2}$ is even and $\mathcal{P} = i\sigma_2$ when $\f{p(p+1)}{2}$ is odd. It is important to note that the projector above is consistent with the reality condition obeyed by the spinors in the type II$^*$ supergravity theory. We would like to stress that this subtle interplay of imaginary R-R fluxes and unusual reality conditions on the spinors is the reason why Euclidean branes preserve supersymmetry in type II$^*$ string theory and supergravity.

\section{Gauged supergravity for spherical branes}
\label{App:sugra}

In this appendix we introduce, case by case, the $(p+2)$-dimensional gauged supergravity theories used to construct the spherical D$p$-brane solutions discussed in the main text. The supergravity theories available in the literature are Lorentzian and we need to analytically continue them to Euclidean signature. After presenting in some detail the construction of the gauged supergravity solutions we perform their uplifts to ten-dimensional type II and/or eleven-dimensional supergravity.

As emphasized in the main text, we start with a maximally supersymmetric gauged supergravity theory in $p+2$ dimensions and perform a consistent truncation, following the method of \cite{Warner:1983vz}, to preserve an $\SO(3)\times \SO(6-p)$ subgroup of the $\SO(9-p)$ gauge group, corresponding to the $R$-symmetry in the dual SYM theory. By analytically continuing the supergravity theory to Euclidean signature we end up with a non-compact $\SO(1,2)\times \SO(6-p) \simeq \SU(1,1)\times \SO(6-p)$ gauge group, in harmony with \eqref{NonCompactR}. We start with the case $p=6$ and work our way down to $p=2$.

\subsection{Spherical D6-branes}
\label{appB:D6}

The supergravity theory appropriate for our construction is the maximal $\SO(3)$ gauged supergravity in eight dimensions, originally constructed in \cite{Salam:1984ft}, analytically continued to Euclidean signature and non-compact gauge group. The uplift of this theory to eleven-dimensional supergravity as well as more general gaugings are discussed in \cite{AlonsoAlberca:2003jq}.

\subsubsection{Maximal eight-dimensional $\SO(3)$ gauged supergravity}

The maximal $\mathcal{N}=2$ ungauged supergravity theory in eight dimensions has $\text{E}_{3(3)} \simeq \SL(3,\mathbf{R})\times \SL(2,\mathbf{R})$ global symmetry under which the bosonic fields of the theory transform. In particular, the 7 scalars parametrize the five-dimensional and two-dimensional coset spaces $\SL(3,\mathbf{R})/\SO(3)$ and $\SL(2,\mathbf{R})/\SO(2)$ and are most conveniently expressed in terms of two matrices $\mathfrak{Z}$ and $\mathfrak{A}$ transforming according to
\begin{equation}
\begin{aligned}
\mathfrak{Z} & \rightarrow G \mathfrak{Z} H\,, \qquad \text{where} \quad G\in \SL(3,\mathbf{R}) \quad \text{and}\quad H \in \SO(3)\,,\\
\mathfrak{A} & \rightarrow K \mathfrak{A} L\,,  \qquad \text{where} \quad K\in \SL(2,\mathbf{R}) \quad \text{and}\quad L \in \SO(2)\,.
\end{aligned}
\end{equation}
The fermionic fields transform under $\SO(3)\times \SO(2)\simeq \SU(2)\times\text{U}(1)$ which acts as the $R$-symmetry of the supergravity theory.\footnote{We are cavalier about the global difference between $\SO(3)$ and $\SU(2)$.} In total, the field content of the ungauged theory consists of the metric $g_{\mu\nu}$, two sixteen-component gravitini $\psi_\mu^a$, six gaugini $\lambda^a_i$, seven scalars ${\mathfrak{Z}_M}^i$ and $\mathfrak{A}_{IJ}$, one three-form tensor field $A_{\mu\nu\rho}$, three two-form tensor fields $A_{\mu\nu}^{M}$, and six one-form vector fields $A_\mu^{MN}$. We use the following index conventions: $\mu,\nu,\rho = 0,\dots 7$ are eight-dimensional spacetime indices; $M,N = 1,2,3$ are $\SL(3,\mathbf{R})$ indices in the fundamental;  $I,J=1,2$ are $\SL(2,\mathbf{R})$ indices in the fundamental; $i,j=1,2,3$ are in the $\mathbf{3}$ and $a,b = 1,2$ are in the $\mathbf{2}$ of $\SU(2)\simeq\SO(3)$, respectively.

To obtain a gauged supergravity theory with a non-trivial potential for the scalars a subgroup of the global symmetry group should be promoted to a local symmetry. This can be done in several inequivalent ways by gauging a subgroup of the global symmetry group. By gauging the maximal compact subgroup $\SO(3)$ in $\SL(3,\mathbf{R})$ one obtains the theory studied by Salam and Sezgin in \cite{Salam:1984ft}. This theory can be obtained by reducing the eleven-dimensional supergravity to eight dimensions on an $\SU(2)$ group manifold. As described in \cite{AlonsoAlberca:2003jq} one can also obtain more general gaugings by reducing the eleven-dimensional supergravity on different group manifolds. One example is a reduction on an $\SU(1,1)$ group manifold resulting in the Lorentzian eight-dimensional $\SO(1,2)\simeq \SU(1,1)$ gauged supergravity. This case can be understood as an analytic continuation of the Salam-Sezgin theory or as a ``non-compactification'' of eleven-dimensional supergravity. However, this $\SU(1,1)$ gauged supergravity theory is still Lorentzian. To obtain the Euclidean action appropriate for constructing the spherical brane solutions of interest we need to combine this analytic continuation of the gauge group with an analytic continuation of the time direction in space-time.

\subsubsection{$\SO(3)$ invariant truncation}

We begin with the $\SO(3)$ gauged supergravity of \cite{Salam:1984ft} and are interested in constructing solutions which preserve the $\SO(3)$ gauge symmetry and have a maximally symmetric seven-dimensional factor in the metric. These requirements eliminate all tensor fields in the supergravity theory except the metric itself. There are two scalars, a ``dilaton'' $\beta$ and an ``axion'' $\chi$, parametrizing an $\SL(2,\mathbf{R})/\SO(2)$ coset, which are not charged under the $\SO(3)$ gauge symmetry.

The Lagrangian for the bosonic fields in this $\SO(3)$ invariant truncation reads\footnote{Our notation is different from the one in \cite{Salam:1984ft}. We have defined $\beta \equiv - 2\phi_{{\rm SS}}$, $\chi \equiv -2B_{{\rm SS}}$, $g \equiv \frac{g_{{\rm SS}}}{2}$, where quantities with an SS subscript are the ones used in \cite{Salam:1984ft}.}   
\be
S = \f{1}{2 \kappa_8^2}\int \star_8 \left\{ R - \f12 \left(|\dd \beta|^2 + \e^{2\beta}|\dd \chi|^2\right) - V \right\}\,.
\ee
The potential is proportional to the gauge coupling constant, $g$, of the supergravity theory and is given by
\be\label{eq:V8d}
V = -\f32 g^2 \e^{\beta}\,.
\ee
It proves convenient to introduce the complex scalar $\tau= \chi+i \e^{-\beta}$ as in \eqref{taudef}, the K\"ahler potential as in \eqref{eq:Kpotdef}, and superpotential as in \eqref{eq:Wp=6}
The potential in \eqref{eq:V8d} can then be written in terms of the superpotential \eqref{eq:Wp=6} using the expression \eqref{eq:defVp=6}.

The supersymmetry variations of this truncation of the supergravity theory can be read off from \cite{AlonsoAlberca:2003jq,Salam:1984ft}. They can be explicitly written as
\bea\label{eq:8dsusyvar}
\delta \psi_\mu &=& \nabla_\mu \epsilon +\f1{8}\e^{\cal K}\partial_\mu (\tau+\bar{\tau})\epsilon +\f{1}{24}\e^{\mathcal{K}}\mathcal{W} \gamma_9\gamma_\mu\epsilon~,\\
\delta \lambda_i &=& \left( \frac{\tau-\bar{\tau}}{2}\partial_\mu\bar{\tau}\gamma_9\gamma^\mu +  \f16 \e^\mathcal{K}D_{\tau}\mathcal{W} \right)\sigma_i\epsilon~,
\eea
where $\gamma_9 = i \gamma^{01\dots 7}$, the spinor $\epsilon^a$ is in the $\mathbf{2}$ of $\SO(3)$, $(\sigma_{i})^{a}_{b}$ are $\SO(3)$ Pauli matrices, and $D_{\tau}$ is the K\"ahler covariant derivative defined below \eqref{eq:defVgenp}.

As described in the main text, we are interested in an analytic continuation of this gravitational theory and its supersymmetry variations into Euclidean signature. This is achieved by changing the signature of the metric as well as replacing the pseudoscalar as follows, $\chi \rightarrow i \chi$. In addition we should treat the complex conjugate of the scalar $\tau$ as an independent scalar. We emphasize this by using the notation $\overline{\mathcal{W}}\rightarrow \widetilde{\mathcal{W}}$ and $\bar{\tau}\rightarrow \tilde{\tau}$.

To find the solution of interest we impose the usual spherically symmetry domain wall ansatz for the metric as in Equation \eqref{eq:genpmetAnsatz}
\be\label{eq:8dmetrappB}
\dd s_8^2 = \dd r^2 + {\cal R}^2 \e^{2A} \dd \Omega_7^2~,
\ee
and assume that the the scalar fields only depend on the radial coordinate $r$.

To solve the supersymmetry variations in \eqref{eq:8dsusyvar} we use a conformal killing spinor on $S^7$ obeying
\be
\nabla^{S^7}_{\mu} \epsilon = \f{i}{2}\gamma_{\mu}\epsilon\,.
\ee
Here $\nabla^{S^7}$ is the covariant derivative on the unit radius $S^7$. Note that this is in harmony with the expected supersymmetry generator for the seven-dimensional SYM theory on $S^7$, see \eqref{spinorderivative}.

The vanishing of the gaugino and gravitino variations then leads to the following differential equations 
\bea\label{BPS6}
\mathcal{K}_{\tilde{\tau}\tau}(\tilde{\tau}^\prime)(\tau^\prime) &=& \f{1}{16}\e^\mathcal{K}\mathcal{W}\widetilde{\mathcal{W}}\,,\\
(A')^2 - \mathcal{R}^{-2}\e^{2A}   &=& \f{1}{144}\e^{2\mathcal{K}}\mathcal{W}\widetilde{\mathcal{W}}\,,
\eea
where a prime denotes differentiation with respect to $r$.  Notice that the equations in \eqref{BPS6} correspond to a degenerate limit of \eqref{lambdaeq}-\eqref{Aeq2} in which we remove the scalar $\lambda$ and set $p=6$.

There is a subtlety when analyzing the BPS equations in this truncation of the eight-dimensional supergravity. There are only two independent equations in \eqref{BPS6} and thus one of the two scalars in the model appears to not be constrained by a differential equation. This problem is fixed by the equations of motion which lead to the following first order differential equation for the scalar $\chi$ 
\begin{equation}\label{eq:chieqn8d}
\chi^\prime = \frac{6}{\mathcal{R}} \e^{-2\mathcal{K}}\e^{-7A}\,.
\end{equation}
We have a first order equation in \eqref{eq:chieqn8d} because the scalar $\chi$ does not appear in the potential $V$ in \eqref{eq:V8d} and the usual second order differential equation has an integral of motion which reduces the order of the equation. The constant coefficient on the right hand side of \eqref{eq:chieqn8d} is the unique value of this integral of motion which makes the BPS equations in \eqref{BPS6} together with \eqref{eq:chieqn8d} compatible with all other equations of motion and with the integrability of the supersymmetry variations in \eqref{eq:8dsusyvar}.

The gauged supergravity solution discussed above can be uplifted to type IIA and eleven-dimensional supergravity and the explicit result is presented in Section ~\ref{subsec:D6}.

\subsection{Spherical D5/NS5-branes}
\label{appB:D5}

To construct a supergravity solution describing spherical D5- or NS5-branes we consider the maximal seven-dimensional $\SO(4)$ gauged supergravity constructed in \cite{Samtleben:2005bp}. We then use the results of \cite{Malek:2015hma} to uplift this seven-dimensional solution to ten-dimensional type IIB supergravity.

\subsubsection{Maximal seven-dimensional $\SO(4)$ gauged supergravity}
\label{subsubsec:7dSO4sugra}

The maximal ${\cal N}=4$ ungauged supergravity theory in seven dimensions has $\text{E}_{4(4)}=\SL(5)$ global symmetry under which the bosonic fields transform. In particular the fourteen scalars span the coset space $\SL(5)/\SO(5)$ and can be parametrized by a matrix ${\mathfrak{Z}}$ that transforms according to
\be\label{eq:calV7ddef}
{\mathfrak{Z}}\to G {\mathfrak{Z}} H\,,\quad \text{where}\quad G\in \text{\SL(5)}\quad \text{and}\quad H\in \text{\SO(5)}\,. 
\ee
In addition to the bosonic fields, the fermions transform under $\SO(5)\simeq\text{USp}(4)$ which acts as the $R$-symmetry group of the supergravity theory. In total the field content of the ungauged theory consists of the metric $g_{\mu\nu}$, four gravitini $\psi_\mu^a$, five two-forms $B_{\mu\nu}^M$, ten vector fields $A_\mu^{MN}$, sixteen gaugini $\chi^{abc}$, and fourteen scalar fields ${\mathfrak{Z}}_M^{\p{M}ab}$. We use the following index conventions: $a,b=1,\dots,4$ denote $\text{USp}(4)$ indices; $M,N=1,\dots,5$ are $\SL(5)$ indices, and $\mu,\nu=0,\dots,6$ are seven-dimensional spacetime indices.

The global symmetries can be promoted to a local symmetry in several inequivalent ways. Gauging the maximal compact subgroup $\text{SO}(5)\subset\text{SL}(5)$ one obtains the well known gauged supergravity theory\cite{Pernici:1984xx}. This theory has a maximally supersymmetric AdS$_7$ vacuum and can also be obtained by performing a consistent truncation of eleven-dimensional supergravity on $S^4$. Further gaugings were discovered in \cite{Pernici:1984zw} and a complete classification was obtained in \cite{Samtleben:2005bp} using the embedding tensor formalism. We are interested in a maximal supergravity with an $\SO(4)$ gauge group which should capture domain wall solutions describing the backreaction of NS5/D5-branes. It was anticipated in \cite{Boonstra:1998mp} that such a supergravity theory should exist and indeed it was explicitly constructed in \cite{Samtleben:2005bp}.\footnote{A half-maximal version of the supergravity theory which can be viewed as a consistent truncation of the maximal theory  was studied in \cite{Salam:1983fa}.}
 
In the maximal $\SO(4)$ gauged theory the $\SL(5)$ representations of the bosonic fields are decomposed into representations of the gauge group. The ten vector fields that transform in the $\obar{\bf 10}$ of $\SL(5)$ transform in $\obar{\bf 6} + \obar{\bf 4}$ of $\SO(4)$ where the $\obar{\bf 6}$ plays the role of the $\SO(4)$ gauge field. Four of the two-forms $B_{\mu\nu}^M$ become massive by combining with the $\obar{\bf 4}$ other vector fields. The fifth two-form is uncharged and is present also in the ${\cal N}=2$ theory \cite{Salam:1983fa}. Finally, the scalars transform in the symmetric traceless of $\SL(5)$, i.e. the ${\bf 14}$, which decomposes into the ${\bf 9}+ {\bf 4}+{\bf 1}$ representation of $\SO(4)$. 

\subsubsection{$\SO(3)$ invariant truncation}

The R-symmetry of the six-dimensional SYM theory on $S^6$ is $\SO(1,2)$ and thus we should find an $\SO(3)$ invariant truncation of the $\SO(4)$ gauged supergravity which we can then analytically continue. Imposing this symmetry on the theory and keeping only fields compatible with a solution preserving the isometries of $S^6$ leads to a consistent truncation of the $\SO(4)$ gauged supergravity consisting of the metric and three real scalar fields. This is in agreement with the field theory discussion in Section~\ref{sec:fieldtheory}. More precisely, the three supergravity scalars should be dual to the Yang-Mills coupling and the two independent operators in the deformation Lagrangian \eqref{symlagrangian}.

The three scalars invariant under the $\SO(3)$ symmetry of interest here are the singlets in the branching rules of the breaking of $\SO(4)$ to $\SO(3)$
\begin{equation}
{\bf 9} \to {\bf 5}\oplus {\bf3}\oplus {\bf1}\,, \qquad {\bf 4} \to {\bf3}\oplus {\bf1}\,, \qquad {\bf 1} \to{\bf1}\,.
\end{equation}
In the notation of \cite{Samtleben:2005bp} the parametrization of the scalar coset element for these three scalars is
\begin{equation}
\mathfrak{Z}= \begin{pmatrix}
\e^{-\phi-x} & 0 & 0 & 0 & 0 \\
0 & \e^{-\phi-x} & 0 & 0 & 0 \\
0 & 0 & \e^{-\phi-x} & 0 & 0 \\
0 & 0 & 0 & \e^{-\phi+3x} & \e^{4\phi}\chi \\
0 & 0 & 0 & 0 & \e^{4\phi} \\
\end{pmatrix}\,.
\end{equation}
Notice that $\chi$ is a pseudoscalar. These three scalars parametrize the following submanifold of the  $\SL(5)/\SO(5)$ scalar coset
\begin{equation}
{\bf R}_+ \times \f{\SL(2,\mathbf{R})}{\SO(2)}\,,
\end{equation}
where ${\bf R}_+$ is parametrized by the combination $\lambda \equiv -\phi-x$ and $\SL(2,\mathbf{R})/\SO(2)$ is parametrized by $\beta\equiv 5\phi-3x$ and $\chi$. The bosonic action for this consistent truncation takes the familiar form \eqref{sugraaction} with $p=5$
\be
S = \f{1}{2 \kappa_7^2}\int \star_7 \left\{R - \f{15}{2}|\dd \lambda|^2 -\f12\left(|\dd\beta|^2+\e^{2\beta}|\dd \chi|^2\right) - V\right\}\,.
\ee
The potential is proportional to the gauge coupling constant $g$ and takes the form
\begin{equation}
	V = \f{g^2}{2}\e^{\beta}(-3\e^{\lambda} - 6\e^{-4\lambda-\beta} + \e^{-9\lambda-2\beta}+\e^{-9\lambda}\chi^2)\,.
\end{equation}
After introducing the scalar $\tau=\chi+i \e^{-\beta}$ one can use the superpotential in \eqref{superpotential} and the K\"ahler potential in \eqref{eq:Kpotdef} to write the potential in terms of the superpotential as in \eqref{eq:defVgenp}.

The supersymmetry variations for this consistent truncation can be obtained from the results in \cite{Samtleben:2005bp}. The vanishing of the spin-$\f12$ variations leads to the equations
\bea
\partial_\mu(\lambda)\gamma^\mu \epsilon^1 = -\f{1}{15}\e^{\mathcal{K}/2}\partial_\lambda\mathcal{W}\epsilon^4~,&& \partial_\mu(\lambda)\gamma^\mu \epsilon^4 = -\f{1}{15}\e^{\mathcal{K}/2}\partial_\lambda\overline{\mathcal{W}} \epsilon^1\,,\\
\partial_\mu \bar{\tau}\gamma^\mu \epsilon^1 = \big(\e^{-\mathcal{K}}\mathcal{K}^{\bar{\tau}\tau}\big)^{1/2}D_\tau\mathcal{W}\epsilon^4~,&&\partial_\mu \tau\gamma^\mu \epsilon^4 = \big(\e^{-\mathcal{K}}\mathcal{K}^{\tau\bar{\tau}}\big)^{1/2}D_{\bar{\tau}}\overline{\mathcal{W}}\epsilon^1\,.
\eea
From the spin-$\f32$ variations we find
\bea
\nabla_\mu\epsilon^1 + \f{i}{8}\e^{\mathcal{K}}\partial_\mu (\tau+\bar{\tau})\, \epsilon^1 &=& -\f{1}{20}\e^{\mathcal{K}/2}\mathcal{W}\gamma_\mu\epsilon^4\,,\\
\nabla_\mu\epsilon^4 - \f{i}{8}\e^{\mathcal{K}}\partial_\mu  (\tau+\bar{\tau})\, \epsilon^4 &=& -\f{1}{20}\e^{\mathcal{K}/2}\overline{\mathcal{W}}\gamma_\mu\epsilon^1\,.
\eea
There are four supersymmetry generators, $\epsilon^a$, in the maximal supergravity theory. However the equations for the pair $(\epsilon^2,\epsilon^3)$ are identical to the ones presented above for $(\epsilon^1,\epsilon^4)$. 

As described in the main text, the analytic continuation to Euclidean signature corresponds, at this level, to the replacement $\chi \rightarrow i \chi$ accompanied by the substitutions $\overline{\mathcal{W}}\rightarrow \widetilde{\mathcal{W}}$ and $\bar{\tau}\rightarrow \tilde{\tau}$. After this analytic continuation we can look for the spherical brane solution by imposing the domain wall metric ansatz as in \eqref{eq:genpmetAnsatz}
\begin{equation}
\dd s_7^2 = \dd r^2 + \mathcal{R}^2 \e^{2A(r)}\dd\Omega_6^2\,,
\end{equation}
and assume that all scalars depend only on the radial coordinate $r$. Furthermore we can assume that $\epsilon_{1,4}$ are conformal Killing spinors on $S^6$
\begin{equation}
	\nabla^{S^6}_\alpha \begin{pmatrix}
	\epsilon_1 \\ \epsilon_4
	\end{pmatrix} = \f12 \gamma_*\gamma_\alpha \begin{pmatrix}
	 \epsilon_1 \\  - \epsilon_4
	\end{pmatrix}\,,
\end{equation}
where $\gamma_* \equiv i \gamma_{123456}$. With this at hand we can derive the system of BPS equations in \eqref{lambdaeq}-\eqref{Aeq2} with $p=5$. Furthermore, combining the spin-$\f12$ and spin-$\f32$ equations we can find an algebraic equation for $A$ as in \eqref{eq:algebraicA} with $p=5$.  

\subsubsection{Uplift to type IIB supergravity}

Any solution of the seven-dimensional  $\SO(4)$ gauged supergravity can be uplifted to the ten-dimensional type IIB supergravity using the uplift formulae of \cite{Malek:2015hma}. When we apply these uplift formulae to the solutions of the BPS equations in \eqref{lambdaeq}-\eqref{Aeq2} with $p=5$ we obtain the spherical NS5-brane solution with the following string frame metric
\begin{equation}
	\dd s^2 = \f{1}{\sqrt{X}}\Big( \e^{-4\lambda} \dd r^2 + \f{X((1-3X)^2-9Y^2)}{g^2Y^2}\dd\Omega_6^2 + \f{1}{g^2}\dd \theta^2 + \f{\sin^2\theta X}{g^2(\sin^2\theta + X\cos^2\theta)}\dd\widetilde{\Omega}_2^2 \Big)~.
\end{equation}
The remaining ten-dimensional fields are given by
\begin{equation}\label{eq:NS5fluxesApp}
	\begin{aligned}
	\e^{2\Phi} &= \f{\e^{-10\lambda}}{X(\sin^2\theta+ X \cos^2\theta)}\,,\\
	C_0 &= iY\e^{5\lambda}\cos\theta\,,\\
	B_2 &= -\f{1}{g^2}\Big( \theta - \f{\sin 2\theta X}{2(\sin^2\theta+X\cos^2\theta)} \Big) \text{vol}_2\,,\\
	C_2 &= -i\f{Y\e^{5\lambda}\sin^3 \theta}{g^2(\sin^2\theta + X\cos^2\theta)}\text{vol}_2\,,
	\end{aligned}
\end{equation}
where $\text{vol}_2$ is the volume element of the $\dd\widetilde{\Omega}^2_2$ metric in \eqref{eq:metdS2}. Integrating the $H$ and $F_3$ flux derived from \eqref{eq:NS5fluxesApp} over the three-dimensional space spanned by $\theta$ and $\dd\widetilde{\Omega}^2_2$ we find that the D5-brane charge is vanishing while the NS5-brane charge is not. This fits nicely with the interpretation of this background as corresponding to spherical NS5-branes.

The spherical D5-brane solution can be obtained from the spherical NS5-brane solution above by acting with the $\SL(2,\mathbf{R})$ global symmetry of the type IIB supergravity. This transformation acts on the supergravity background fields as follows 
\begin{equation}\label{eq:SL2RApp}
	\tau_{\rm IIB} \mapsto \f{a \tau_{\rm IIB} + b}{c\tau_{\rm IIB} + d}~ ,\qquad \begin{bmatrix}
	C_2 \\ B_2
	\end{bmatrix} \mapsto \begin{bmatrix}
	a & b \\ c & d 
	\end{bmatrix} \begin{bmatrix}
	C_2 \\ B_2
	\end{bmatrix}~ , \qquad \text{where} \quad \begin{bmatrix}
	a & b \\ c & d 
	\end{bmatrix} \in \SL(2,\mathbf{R})\,.
\end{equation}
Here $\tau_{\rm IIB} \equiv C_0 + i\e^{-\Phi}$ is the axion-dilaton field and the Einstein frame metric remains unchanged. Applying this transformation to the background in \eqref{eq:NS5fluxesApp} with $a=d=0$ and $b=-c=1$, yields the spherical D5-brane solution in type IIB supergravity. The fluxes of this D5-brane solution are the same as the ones in \eqref{eq:NSNSRRgenp} with $p=5$. In particular the $H$ flux integral over $\dd\theta$ and $\dd\widetilde{\Omega}^2_2$ vanishes while the R-R flux integral over this space does not. For other values of the $\SL(2,\mathbf{R})$ parameters in \eqref{eq:SL2RApp} we obtain more general solutions which should describe $(p,q)$-fivebranes wrapped on $S^6$.

\subsection{Spherical D4-branes}
\label{appB:D4}

If we are to follow the pattern of gauged supergravity theories used to construct spherical D$p$-brane solutions we should use a maximal six-dimensional $\SO(5)$ gauged supergravity. This theory is not so well studied in the literature and the only analysis we are aware of is the one in \cite{Cowdall:1998rs} where the author constructed the six-dimensional theory through a dimensional reduction of the maximal seven-dimensional $\SO(5)$ gauged supergravity on a circle. We will thus describe the spherical D$4$-brane background as a solution of this maximal seven-dimensional supergravity theory. The $\SO(5)$ gauged supergravity has a maximally supersymmetric AdS$_7$ solution dual to the conformal vacuum of the $(2,0)$ six-dimensional SCFT which fits well with the field theory expectation, discussed in Section~\ref{sec:fieldtheory}, that the five-dimensional maximal SYM theory on $S^5$ flows in the UV to the six-dimensional $(2,0)$ theory on $S^5\times S^1$.

The maximal $\SO(5)$ gauged supergravity in seven dimensions was constructed in  \cite{Pernici:1984xx} and can be obtained as a consistent truncation of eleven-dimensional supergravity on $S^4$. Any solution of the seven-dimensional theory can be uplifted to eleven dimensions using the uplift formulae of \cite{Nastase:2000tu}. The field content of the theory is the same as for the $\SO(4)$ gauged supergravity discussed in Appendix~\ref{subsubsec:7dSO4sugra}. The difference comes from the gauging which in this case is $\SO(5)$. This gauging of course fits in the general classification of \cite{Samtleben:2005bp}, whose conventions we use, and affects the details of the Lagrangian of the theory and thus the space of solutions.

\subsubsection{$\SO(3)$ invariant truncation}

The R-symmetry breaking pattern discussed around \eqref{NonCompactR} dictates that we should look for the spherical D4-brane solutions in an $\SO(3)\times \SO(2)$ invariant truncation of the $\SO(5)$ gauged supergravity. This, combined with the requirement that the solution should have the isometries of $S^5\times S^1$, leads to a consistent truncation which consists of the metric, a single real scalar field, $x$, and a single $\SO(2)$ gauged field, $A$.\footnote{The $\SO(2)$ gauge field generator can be thought of as the 45 component of the $5\times 5$ matrix generator of the $\SO(5)$ gauge field.} The scalar coset matrix \eqref{eq:calV7ddef} for this truncation is diagonal and reads
\begin{equation}
\mathfrak{Z}= {\rm diag}(\e^{-x},\e^{-x},\e^{-x},\e^{3x/2},\e^{3x/2})\,.
\end{equation}
The bosonic action can be obtained from \cite{Samtleben:2005bp} and reads
\begin{equation}
	S = \f{1}{2\kappa_7^2}\int \star_7 \left\{R_7 - \f{15}{2}|\dd x|^2 -\f12 \e^{6x} F_{\mu\nu}F^{\mu\nu} - V_7\right\}\,,
\end{equation}
where $F=\dd \mathcal{A}$ is the gauge field strength of $\mathcal{A}$ and the potential is proportional to the gauge coupling constant $g$,
\begin{equation}
	V_7 = -\f{3}{2}g^2 \e^{-x}(4+\e^{5x})\,.
\end{equation}

We can now dimensionally reduce this theory on $S^1$ to a six-dimensional gravitational theory.\footnote{It should also be possible to construct this six-dimensional theory as a consistent truncation of the six-dimensional maximal gauged supergravity studied in \cite{Cowdall:1998rs}.} To this end we use the following metric and gauge field ansatz 
\begin{equation}\label{eq:metAD4app}
	\dd s_7^2 = \e^{-\phi} \dd s_6^2 + \e^{4\phi}\dd\omega^2\,, \qquad \mathcal{A} = \chi \dd\omega\,.
\end{equation}
The scalar fields $\phi$ and $\chi$ depend only on coordinates of the six-dimensional space with metric $\dd s_6^2$. To conform with the notation used throughout this work it is convenient to define the following combination of these two scalars
\begin{equation}
\beta\equiv 3x - 2\phi\,, \qquad \text{and} \qquad \lambda \equiv x + \phi\,.
\end{equation}
The six-dimensional Lagrangian of the dimensionally reduced theory then reads
\begin{equation}
	S = \f{1}{2\kappa_6^2}\int\star_6 \left\{R - 3|\dd \lambda|^2 -\f12 \Big( |\dd \beta|^2 + \e^{2\beta}|\dd \chi|^2 \Big) - V\right\}\,,
\end{equation}
where $R$ is the Ricci scalar for the metric $\dd s_6^2$ and the six-dimensional potential is
\begin{equation}\label{eq:V6def}
	V = -\f{3}{2}g^2 \e^{-\lambda}(4+\e^{\beta+2\lambda})~.
\end{equation}

The derivation of the BPS equations now follows a familiar pattern. We work with the supersymmetry variations of the seven-dimensional maximal supergravity theory as given in \cite{Samtleben:2005bp}. To present them succinctly we define the scalar $\tau=\chi+i\e^{-\beta}$ and the superpotential as in \eqref{superpotential} with $p=4$. Using the K\"ahler potential in \eqref{eq:Kpotdef} one can then show that the six-dimensional potential in \eqref{eq:V6def} can be written in the general form \eqref{eq:defVgenp} with $p=4$. Combining the gaugino and gravitino variations of \cite{Samtleben:2005bp} we find
\bea\label{eq:susyvarD4App}
\partial_\mu\lambda \gamma^\mu\epsilon &=& \f16e^{\mathcal{K}/2} \partial_\lambda  \mathcal{W} \epsilon~,\\
\partial_\mu \tau \gamma^\mu &=& (\e^{-\mathcal{K}}\mathcal{K}^{\bar{\tau}\tau}\big)^{1/2}D_\tau \mathcal{W} \epsilon~,\\
\nabla_\mu\epsilon +\f{i}{8}  \e^{\mathcal{K}}\partial_\mu (\tau+\bar{\tau}) \,\epsilon &=&  - \f{1}{16} \e^{\mathcal{K}/2}\mathcal{W} \gamma_\mu \epsilon~.
\eea

Now we can perform the familiar analytic continuation to Euclidean signature by treating $\tau$ and $\bar{\tau} \to \tilde{\tau}$ as independent scalars and performing the substitution $\chi \to i\chi$. For the metric we use the usual spherical domain wall ansatz
\begin{equation}
\dd s_6^2 = \dd r^2 + {\cal R}^2\e^{2A}\dd \Omega_5^2\,,
\end{equation}
and assume that all scalar fields depend only on $r$. The supersymmetry parameter $\epsilon$ is a conformal Killing spinor on $S^5$ obeying
\begin{equation}
	\nabla_\mu^{S^5}\epsilon = \f{i}{2}\gamma_\mu \epsilon\,.
\end{equation}
We can plug this in the supersymmetry variations \eqref{eq:susyvarD4App} and derive the system of BPS equations in \eqref{lambdaeq}-\eqref{Aeq2} and the algebraic equation in \eqref{eq:algebraicA} with $p=4$. 

\subsubsection{Uplift to eleven-dimensional and type IIA supergravity}

The solution of the maximal seven-dimensional $\SO(5)$ gauged supergravity described above can be uplifted to eleven dimensional M$^*$ theory using the uplift formulae presented in \cite{Nastase:1999cb,Nastase:1999kf}. Using the functions  $P$ and $Q$ defined in \eqref{eq:PQdef} we can write the eleven-dimensional metric as
\begin{equation}
\begin{aligned}
\dd s_{11}^2 =& g_s^{-2/3}\e^{-\lambda}P^{-1/3}\, \dd s_6^2 + g_s^{4/3}\e^{4\lambda}Q^{-1}\, P^{2/3}\,\bigg( \dd x_{11} + \frac{i Y Q\sin^2\theta}{g_sg \e^{2\lambda}X}\dd \zeta \bigg)^2\\
&\quad+ \frac{g_s^{-2/3}P^{-1/3}}{g^2}\Big( \dd\theta^2 + P \,\cos^2\theta\, \dd\tilde{\Omega}_2^2 + Q\,\sin^2\theta \,\dd \zeta^2 \Big)\,,
\end{aligned}
\end{equation}
where $x_{11}$ should be taken as pure imaginary and therefore spans a timelike $\U(1)$ whereas $\zeta$ spans a spacelike $\U(1)$.
The three-form gauge field is
\begin{equation}
A_{(3)} = \frac{P\cos^3\theta}{g^2X}\left(-\f{i}{g_sg} \dd \zeta + \e^{2\lambda}Y \dd x_{11}\right)\wedge \vol_2~,
\end{equation}
where $\vol_2$ is the volume form of the two dimensional de Sitter space in \eqref{eq:metdS2}.

This eleven-dimensional solution can be dimensionally reduced to ten-dimensional type IIA$^*$ supergravity along the timelike $\U(1)$ spanned by $x_{11}$ using the formulae in \eqref{eq:KK11to10}. The result is a type IIA background of the form described in Section~\ref{subsec:uplift} with $p=4$. Notice that we introduced $g_s$ by hand in the above to conform with the notation in Section~\ref{subsec:uplift} and the rest of the paper.

\subsection{Spherical D2-branes}
\label{App:D2}

To construct spherical D2-branes we employ the four-dimensional maximal $\text{ISO}(7)$ gauged supergravity as presented in \cite{Guarino:2015qaa}. We construct spherical brane solutions to this theory which can then be uplifted to both type IIA and eleven-dimensional supergravity. 

\subsubsection{ISO$(7)$ gauged supergravity in four dimensions}

The maximal ungauged supergravity in four dimensions has $\text{E}_{7(7)}$ global symmetry under which the bosonic fields transform. In particular, the scalars parametrize the 70-dimensional coset space $\text{E}_{7(7)}/\SU(8)$ with coset element $\mathfrak{Z}$ that transforms according to
\begin{equation}
	\mathfrak{Z}\rightarrow G\mathfrak{Z}H\,, \qquad \text{where} \quad G\in \text{E}_{7(7)} \quad \text{and} \quad H \in \SU(8)\,.
\end{equation}
In addition to the bosonic fields, the fermions transform under $\SU(8)$ which acts as the $R$-symmetry of the supergravity theory. In total, the field content of the ungauged theory consists of the metric $g_{\mu\nu}$, eight gravitini $\psi_\mu^{i}$, $56$ gaugini $\chi^{ijk}$, $28$ gauge fields $A_\mu^{M}$ and $70$ scalars $\mathfrak{Z}_{M}^{ij}$. Here we use the following index conventions: $\mu,\nu = 0,\dots,3$ are four-dimensional spacetime indices, $M=1,\dots, 56$ are $\text{E}_{7(7)}$ indices and $i,j = 1,\dots, 8$ are $\SU(8)$ indices.

A subgroup of the global symmetry group of the supergravity theory can be promoted to a gauge group in several inequivalent ways. The well-known  $\SO(8)$ gauged supergravity theory described in \cite{deWit:1982bul} is obtained in this way and is relevant for the low-energy dynamics of a system of coincident M2-branes since it arises as a consistent truncation of the eleven-dimensional supergravity on $S^7$. Here we are however interested in D2-branes and thus should use an $\text{ISO}(7)$ gauged four-dimensional supergravity. There are two inequivalent ways to find a four-dimensional maximal supergravity theory with an $\text{ISO}(7)$ gauge group. The ``electrically gauged'' theory was constructed in \cite{Hull:1984yy} (see also \cite{Hull:1988jw}) and is the one that admits an uplift to type IIA supergravity with vanishing Romans mass. This will be the theory we focus on for our analysis. The other inequivalent gauging is described in detail in \cite{Guarino:2015qaa} and is relevant for compactifications of the massive type IIA supergravity on $S^6$ \cite{Guarino:2015jca}.

\subsubsection{$\SO(3)$ invariant truncation}

We use the results of \cite{Guarino:2015qaa} and focus on the electric gauging with $m=0$ relevant for type IIA supergravity with vanishing Romans mass. We want to study a solution that preserves $\SO(4)\times\SO(3)$ gauge symmetry and has maximally symmetric three-dimensional factor in the metric. This truncation eliminates most scalars and all tensor fields except the metric. A larger truncation of this four-dimensional supergravity which imposes only $\SO(4)$ symmetry was studied in Section 5 of \cite{Guarino:2015qaa}. The $\SO(4)\times\SO(3)$ truncation of interest here can be obtained from that larger truncation by setting one of the pseudoscalars in \cite{Guarino:2015qaa} to zero. To comply with the notation used in the main text we make the following change of notation with respect to \cite{Guarino:2015qaa}
\be
\lambda\equiv 2\varphi_\text{GV}\,,\qquad \beta\equiv \phi_\text{GV}\,,\qquad \chi \equiv \rho_\text{GV}\,,\qquad \chi_\text{GV}= 0\,,
\ee
where the subscript GV refers to the quantities used in Section 5 of \cite{Guarino:2015qaa}.

The bosonic Lagrangian of this supergravity truncation can be read off from \cite{Guarino:2015qaa}  
\be\label{eq:Lag4dISO7}
{\cal L} = \star_4\left\{ R - \f34 |\dd \lambda|^2 - \f12\left(|\dd \beta|^2+\e^{2\beta}|\dd \chi|^2\right) - V \right\}\,,
\ee
where the potential $V$ is given by
\be\label{eq:pot4dISO7}
V = -\f12 g^2\e^{-\beta}\left[24\e^{\lambda/2+\beta}+8\e^{2\beta}+3\e^{\lambda}(1+\chi^2\e^{2\beta})\right]\,.
\ee
and as usual $g$ is the gauge coupling. Notice that this Lagrangian is of the general form in \eqref{sugraaction} with $p=2$.

We are not aware of a reference in the literature where the explicit fermionic supersymmetry variations for this $\ISO(7)$ gauged supergravity were presented. However the authors of \cite{Guarino:2015qaa} write down an explicit superpotential for our truncation and, after defining $\tau=\chi+i\e^{-\beta}$, it can be readily checked that it coincides with the one in in \eqref{superpotential} with $p=2$. Using the K\"ahler potential in \eqref{eq:Kpotdef} one can then show that the potential in \eqref{eq:pot4dISO7} can be written in the general form \eqref{eq:defVgenp} with $p=2$. In addition one can show explicitly that the system of BPS equations in \eqref{lambdaeq}-\eqref{Aeq2} and the algebraic equation \eqref{eq:algebraicA} (with $p=2$) imply the equations of motion derived from the Lagrangian in \eqref{eq:Lag4dISO7}.\footnote{We perform our usual analytic continuation to Euclidean signature and take a spherically symmetric ansatz for the four-dimensional metric and scalar fields.} We consider these results as sufficient evidence that any solution to the system of equations in \eqref{lambdaeq}-\eqref{Aeq2} describes a supersymmetric solution of the four-dimensional $\ISO(7)$ electrically gauged supergravity theory.

Finally we point out that one can use the uplift formulae provided in \cite{Guarino:2015jca} to uplift any solution of the four-dimensional $\ISO(7)$ gauged supergravity to a solution of type IIA supergravity. For the $\SO(4)\times \SO(3)$ truncation described above this uplifted ten-dimensional background has the form presented in Section \ref{subsec:uplift}. The ten-dimensional solution has vanishing Romans mass and thus can be further uplifted to a solution of eleven-dimensional supergravity using the formulae in \eqref{eq:KK11to10}.

\bibliographystyle{utphys}
\bibliography{refs}

\end{document}